\documentclass[fleqn,useAMS,usenatbib]{mnras}

\usepackage[T1]{fontenc}
\usepackage{ae,aecompl}

\usepackage{graphicx}
\usepackage{amsmath}
\usepackage{subfig}
\usepackage{mathrsfs} %for \mathscr
\usepackage[squaren]{SIunits}

\newcommand{\eqn}[1]{Eqn.~(\ref{#1})}
\newcommand{\ssec}[1]{Sec.~\ref{#1}}
\newcommand{\fig}[1]{Fig.~\ref{#1}}
\newcommand{\app}[1]{App.~\ref{#1}}
\newcommand{\mean}[1]{\left\langle #1\right\rangle}

\title[The GI $\gamma$-ray Background and Dark Matter]{The Galactic Isotropic $\gamma$-ray Background and Implications for Dark Matter}
\author[S.~S.~Campbell et al.]{
Sheldon~S.~Campbell\thanks{E-mail: sheldoc@uci.edu},
Anna~Kwa,
and Manoj~Kaplinghat
\\
Center for Cosmology, Department of Physics and Astronomy, University of California, Irvine, CA 92697, USA}

\date{Accepted XXX. Received YYY; in original form ZZZ}

\pubyear{2017}

\begin{document}
\label{firstpage}
\pagerange{\pageref{firstpage}--\pageref{lastpage}}
\maketitle

\begin{abstract}
We present an analysis of the radial angular profile of the galacto-isotropic (GI) $\gamma$-ray flux---\,the statistically uniform flux in circular annuli about the Galactic center. Two different approaches are used to measure the GI flux profile in 85 months of Fermi-LAT data: the BDS statistic method which identifies spatial correlations, and a new Poisson ordered-pixel method which identifies non-Poisson contributions. Both methods produce similar GI flux profiles. The GI flux profile is well-described by an existing model of bremsstrahlung, $\pi^0$ production, inverse Compton scattering, and the isotropic background. Discrepancies with data in our full-sky model are not present in the GI component, and are therefore due to mis-modeling of the non-GI emission. Dark matter annihilation constraints based solely on the observed GI profile are close to the thermal WIMP cross section below 100 GeV, for fixed models of the dark matter density profile and astrophysical $\gamma$-ray foregrounds. Refined measurements of the GI profile are expected to improve these constraints by a factor of a few.
\end{abstract}

\begin{keywords}
methods: data analysis -- methods: statistical -- gamma-rays: diffuse background -- gamma-rays: ISM -- dark matter
\end{keywords}

\section{Introduction}

Since its launch in 2008, the Fermi Large Area Telescope (Fermi-LAT) on board the Fermi $\gamma$-ray Space Telescope has significantly focused our understanding of astrophysical $\gamma$ rays of energies $\sim\!\!0.3$--$800$ GeV and their sources. In addition to emission from many point sources \cite{Acero:2015hja}, there is also diffuse emission produced predominantly by unresolved point sources and interactions of energetic cosmic rays with gas, dust, and background light \citep{FermiLAT:2012aa,DiMauro:2015tfa}. 

However, much remains to be understood about the $\gamma$-ray sky. The spatial distribution of gas and starlight can be accurately mapped at angular resolutions much more refined than can currently be resolved in $\gamma$-rays; therefore, it makes sense to attempt to use spatial templates to identify components of the diffuse $\gamma$ rays that originate from inverse Compton scattering off starlight, or originate from interactions of cosmics with gas clouds. Unfortunately, even these $\gamma$-ray spatial templates are sensitive to uncertainties in the distribution of cosmic-ray sources and properties of cosmic-ray propagation in the interstellar medium (ISM). As such, while multiple models exist that can individually agree with many features, present analyses of diffuse $\gamma$-rays have not yet produced a clearly favored model that accounts for all properties of observed $\gamma$-rays \citep{FermiLAT:2012aa}.

Without a fundamental parametrised model of $\gamma$ ray sources, it is unclear how best to improve modeling. At the same time, it is clear that spatial template fitting, while being a tractable analysis, does not use all information in the Fermi-LAT data. Therefore, new observables involving $\gamma$-rays have the potential to provide additional constraints on the sources. 

There are many examples of analyses using alternative observables with the $\gamma$-ray data to further constrain properties of known sources or constrain the presence of exotic sources. Analyses involving the diffuse isotropic background (a largely structureless component) are sensitive to the population of distant, extragalactic, unresolved point sources, as well as the isotropic component of Galactic emission \citep{Ackermann:2014usa,Ackermann:2015tah}. Structured components have been studied via the angular correlation functions, angular power spectrum \citep{Fornasa:2016ohl,Ando:2017alx}, or wavelets \citep{Bartels:2015aea,McDermott:2015ydv}, among others. The presence of random, non-Poissonian sources has been studied using pixel statistics \citep{Malyshev:2011zi,Lee:2014mza,Lee:2015fea,Zechlin:2015wdz,Zechlin:2016pme}. Cross-correlating the data with galaxy catalogs has the promise to identify components of the data that come from particular extragalactic emitters with characteristic spectra, such as blazars or star-forming galaxies \citep{Xia:2015wka}. Cross-correlating with weak lensing or cosmic shear selects emission from the most massive structures in the Universe \citep{Camera:2014rja,Shirasaki:2016kol,Troster:2016sgf}.

By restricting an analysis to particular structured or unstructured components, the signal-to-noise of sources contributing to them can be increased and the models of those sources can be refined to agree with each new observable.

With this motivation, we analyze the galacto-isotropic (GI) component of $\gamma$ rays---spatially structureless emission over circular annuli centered on the Galactic center, independently determined at each radius. A useful diagnostic of any full-sky $\gamma$ ray model is checking the fidelity of its GI profile with the observed profile. Such a radial profile about the Galactic center is also the dominant feature expected from the distribution of dark matter annihilation, as observed from our location inside but offset from the center of a large approximately-spherical dark matter halo. The observed $\gamma$-ray skymap is strongly non-GI, with most of the emission contained along the Galactic plane or from bright point sources. Extracting and understanding the GI component is a robust technique for increasing the signal-to-noise for exotic GI components, including dark matter annihilation.

There is extensive indirect evidence for the presence of dark matter from its gravitational effects on astronomical kinematics and cosmological dynamics \citep{Bertone:2004pz}. One of the simplest models of particle dark matter is a thermally produced big-bang relic of a weakly interacting massive particle (WIMP). In this scenario, dark matter interactions with the big-bang thermal bath keep it in thermal equilibrium. If these interactions lead to dark matter self-annihilation with a strength on the order of the weak nuclear force, the present observed abundance of dark matter is naturally explained. Dark matter of this nature will continue to annihilate predominantly in the densest regions of the Universe, and if so, these annihilations must produce $\gamma$-rays. However, the produced annihilation spectrum depends on how dark matter is coupled to the standard model. Thus, an observed annihilation spectrum from astrophysical dark matter would revolutionize our understanding of the nature of dark matter and provide vital clues about particle physics beyond the standard model.

Searches for gamma rays from dark matter annihilation have been carried out by many analyses. Most of these analyses focus on small regions of the sky with lines-of-sight known to contain a nearby overdensity of dark matter. The brightest such region of WIMP annihilation is the area near the galactic center, but the astrophysical contributions in this region are also very bright. Modeling this emission contributes significant uncertainty to any anomalous flux \citep{Abazajian:2014fta,Zhou:2014lva,Calore:2014xka,TheFermi-LAT:2015kwa}. Nonetheless, a robust significant excess of $\gamma$-rays is observed and has been extensively studied \citep{Goodenough:2009gk,Vitale:2009hr,Cumberbatch:2010ii,Hooper:2010mq,Abazajian:2010zy,Hooper:2011ti,Abazajian:2012pn,Hooper:2013rwa,Gordon:2013vta,Mirabal:2013rba,Macias:2013vya,Abazajian:2014fta,Daylan:2014rsa,Lacroix:2014eea,Yuan:2014rca,Carlson:2014cwa,Petrovic:2014uda,Zhou:2014lva,Abazajian:2014hsa,Petrovic:2014xra,Yuan:2014yda,Calore:2014nla,Kaplinghat:2015gha,OLeary:2015qpx,Bartels:2015aea,Cholis:2015dea,Lee:2015fea,Brandt:2015ula,Gaggero:2015nsa,TheFermi-LAT:2015kwa,Lacroix:2015wfx,OLeary:2016cwz,Carlson:2016iis,Linden:2016rcf,Horiuchi:2016zwu,Macias:2016nev,Karwin:2016tsw,Haggard:2017lyq,TheFermi-LAT:2017vmf,Fermi-LAT:2017yoi,Ploeg:2017vai}.

Currently, the highest, most robust signal-to-noise for dark matter annihilation is from stacking analyses of Milky Way dwarf satellites, which are dark matter dominated and $\gamma$-ray quiet \citep{Ahnen:2016qkx}. The dwarf satellite fields-of-view contain a very small solid angle of the sky, but the dark matter halo and its substructure are everywhere. This leads one to wonder if a larger signal-to-noise can be attained (without sacrificing robustness) by analyzing a much larger region of the sky.

One such strategy is to search for energy-dependent modulations in the gamma-ray intensity or anisotropy at high galactic latitudes. However, the robustness of constraints from such methods currently suffer from uncertainties in the small-scale structure of WIMPs below the resolution of cosmological simulations. In addition, coherent transient structures--for example, streams and shells--are expected relics from the formation history of our galaxy (which is currently unknown), but the magnitude of their contributions to the mean intensity or anisotropy of dark matter annihilation cannot be known. As such, conservative constraints based on limiting the presence of coherent small-scale structure are not very competitive, though they still remain interesting as potential discovery channels.

Dark matter analyses based on the isotropic $\gamma$-ray background (IGRB) are also very sensitive to the nature of small-scale structure. In some realistic models, the contribution of Galactic and extragalactic annihilations to the IGRB can be of similar intensity \citep{Fornasa:2012gu,Ando:2013ff}, such that uncertainties in both components may be important. Also, the signal-to-noise of the IGRB analysis suffers because the isotropic component only includes the dimmest contribution of the Galactic annihilation (corresponding to the intensity from near the anti-Galactic center for a spherical halo).

We will show that an analysis of the GI $\gamma$ rays overcomes many of these difficulties. \citet{Baxter:2011rc} first proposed considering the GI component to robustly increase signal-to-noise for dark matter annihilation in Fermi-LAT data. They used 28 months of data to estimate the GI component at energies $\unit{1}{\giga\electronvolt}<E<\unit{100}{\giga\electronvolt}$ in $1^\circ$ pixels, pioneering the use of the Brock-Dechert-Scheinkman (BDS) statistic for this application. The BDS test checks for spatial correlations in a data string, determining a p-value for the hypothesis that the string of data is isotropic and identically distributed (IID) (see \ssec{ssec:bds} for more details). The GI component that was measured was found to be most constraining of dark matter at positions about $15^\circ$ from the Galactic center, producing modest constraints on the dark matter annihilation cross section.

We update and expand this work to analyze the GI component using 85 months of Fermi-LAT data over the energy range $\unit{2}{\giga\electronvolt}<E<\unit{300}{\giga\electronvolt}$. Additional data improves the contrast of structure. Resolving additional structure lowers the measured GI flux, improving source constraints and increasing sensitivity for exotic searches. In addition, we carry out the following enhancements to the analysis.
\begin{enumerate}
\item We significantly improve the masking of structure by interleaving pixels above and below the Galactic plane.
\item We find that different pixel sizes are appropriate for the BDS analysis at different radii from the Galactic center. Thus, we carry out the BDS analysis at 3 angular resolutions, using $1^\circ$, $0\fdg5$, and $0\fdg2$ diameter pixels. At higher resolution, the GI component can be resolved closer to the Galactic center and smaller structures are identified and masked, but at large radii the counts per pixel become too small. In that case, the quantization of flux and high frequency of empty pixels makes the algorithm unstable. 
\item We consider spectral information by partitioning the data into 4 energy bins and considering the GI component of each bin.
\item For illustration, we compare the measured GI components to a typical self-consistent model of the diffuse $\gamma$-ray sky using spatial templates of inverse Compton scattering \citep{Stecker:1977ua,Moskalenko:1998gw}, bremsstrahlung \citep{Koch:1959zz,Stecker:1975tz,Strong:1998fr}, and $\pi^0$ meson production \citep{Stecker:1971ivh,Stecker:1973sx,1986A&A...157..223D,1986ApJ...307...47D,Moskalenko:1997gh}. This model is used to illustrate preliminary GI constraints of dark matter annihilation.
\end{enumerate}

As alluded to in point~(2), the BDS algorithm becomes unstable under two conditions. If too near the Galactic center, the number of the pixels becomes too small for the BDS test to be statistically reliable. If too far from the Galactic center, the data can become too sparse at high Galactic latitudes such that the data string can no longer be assumed to be drawn from a continuous random variable, and the BDS test is no longer valid. The sparseness of the data prevents some annuli from being analyzed at the angular resolution and energy resolution of the instrument.

To measure the GI component under these sparse conditions, we introduce a more robust statistical method. Rather than masking structure, this new algorithm is based on masking hot pixels that spoil a uniform Poisson distribution of the annulus. This method hypothesizes the existence of a uniform Poisson flux, and determines its maximum possible magnitude that is consistent with the data. This uniform Poisson flux is identified as the GI flux for that annulus.

This Poisson method is found to extract a nearly identical GI component as the BDS test where it is valid. This makes sense if bright non-Poissianties are predominantly due to structured sources such as extended gas clouds and bright point sources. Since the results of the two methods are so similar, we only introduce the Poisson method of measuring the GI component, but use the BDS method for all of our main results.

The details and results are organized as follows. In \ssec{sec:data}, we describe the data used for this analysis, and introduce the pixelization scheme--a galacto-isotropic tiling. The BDS test and Poisson test are briefly explained in \ssec{sec:measurement} and we present the resulting GI profile of the $\gamma$-ray sky. A model of this emission is given in \ssec{sec:model}, resulting dark matter constraints are shown in \ssec{sec:dm}, and a summary of the results is provided in \ssec{sec:discussion}.

\section{Fermi-LAT Data and Galacto-Isotropic Tiling}
\label{sec:data}

We utilized approximately 85 months of Pass 8 public data from the Fermi Large Area Telescope (LAT) instrument taken between August 2008 and September 2015. We used both front- and back-converting \texttt{SOURCE}-class photon events and the corresponding Pass 8 instrument response functions. Events range from 2--300 GeV and are binned into smaller energy bins during the later analysis. A maximum zenith angle cut of 90$^\circ$ is applied to avoid contamination from the Earth's limb. The data reduction, cuts, and exposure map calculation were all performed using the Fermi Science Tools software package\footnote{Fermi Science Tools v10r0p5 \url{https://fermi.gsfc.nasa.gov/ssc/data/analysis/software}}.

To best facilitate the determination of the GI profile, we pixelize the data in a GI tiling. In this paper, we use the following method. When the pixel diameter $\theta$ is specified, we begin with an inner circular pixel of diameter $\theta$ centered on the Galactic center, and then draw concentric annuli of radial thickness $\theta$ beyond that.

In this work, the pixel edges are aligned along the geodesic from the Galactic center to the Galactic North Pole. The pixel azimuthal width in each annulus is rounded to the value nearest to $\theta$ that allows an integer number of equivalent pixels in the annulus. The GI tiling flux map within $60^\circ$ from the Galactic center is presented with $1^\circ$ pixels in the left panel of \fig{fig:GItiling1}. GI maps with $0\fdg5$ and $0\fdg2$ pixels are provided in \app{app:highprecgi}. The smaller pixel size of $0\fdg2$ is approximately the size of the Fermi-LAT's point spread function (PSF) at the higher energy range of this data (see \app{app:psf}). We choose the lower-bound energy of 2 GeV because the width of the PSF increases significantly at lower energies.

\begin{figure*}
  \includegraphics[width=0.85\linewidth]{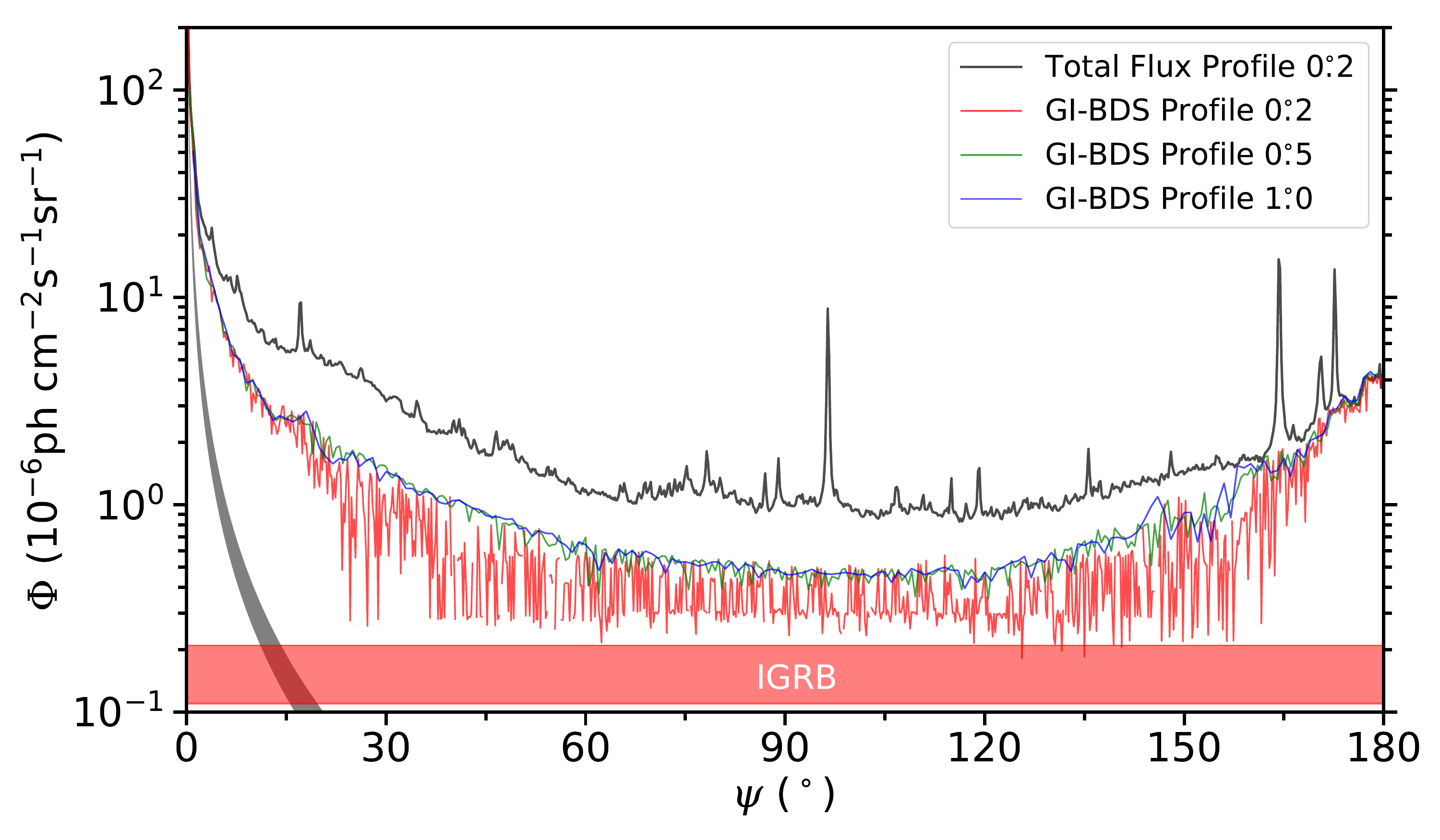}
  \caption{\label{fig:fluxplot}The galacto-isotropic (GI) flux determined using the Brock-Dechert-Scheinkman (BDS) test for the presence of one-dimensional spatial correlations along annuli. The black line indicates the total $\gamma$-ray mean radial flux profile about the Galactic center at $0\fdg2$ resolution. The lower lines show the BDS-determined GI flux (with North-South interleaving) in each annulus for different pixel resolutions. Each resolution has its own range where the algorithm is stable and applicable; for instance, the $0\fdg2$ resolution is only stable near the Galactic center or anti-center. Also indicated are the range of modeled fluxes of the isotropic background (labeled IGRB), and a flux shown in gray near the Galactic center that is consistent with the GeV Galactic center excess (see \ssec{sec:dm} for details).}
\end{figure*}

The exposure-weighted mean observed radial flux profile is shown by the black line in \fig{fig:fluxplot} for the $0\fdg2$ pixel resolution.

\section{The Galacto-Isotropic Profile}
\label{sec:measurement}
We now describe two different methods for measuring the GI profile of astrophysical $\gamma$ rays at Earth. The similarities between these methods is that they both start with similar assumptions about the GI and non-GI component fluxes.  The GI flux of an annulus is assumed to be a constant spectrum of uniform intensity, which may arise from a combination of different $\gamma$-ray sources. The non-GI flux is presumed to be localized contaminations that make pixels too hot to reasonably have the same flux as the other pixels in its annulus. Such contaminations include bright point sources or cosmic-ray interactions with more extended gas and dust clouds (especially along the Galactic disk). The assumption of localized contaminations carries the intended implication that a significant fraction of the annulus pixels are uncontaminated. If observations are inconsistent with this assumption, the underlying GI flux may remain either undetected or poorly resolved by our proposed detection methods.

The first method focuses on the fact that the observed flux in the uncontaminated pixels must be consistent with being drawn from independent and identically distributed (IID) probability distribution functions (PDF). The BDS test provides a p-value on the hypothesis that a string of data (the fluxes of our pixels in this case) are drawn from an IID distribution. If a set of pixels fails this test, we can assume that the brightest pixels must be contaminated, and continue to remove them until the remaining pixels are IID to sufficient confidence. The remaining pixels are then consistent with being uncontaminated, and their median flux provides an estimator for the GI flux of that annulus. This test is only concerned that the pixels of the GI flux be spatially uncorrelated, but is unconcerned with the PDF of the pixels, which we know must be Poisson distributed. Deviations from a Poisson distribution would be an indication of further contamination by a non-GI component (which happens to be spatially uncorrelated).

The second method is more focused on the fact that uncontaminated GI-flux pixels are drawn independently from a Poisson distribution. Since we expect the dimmest pixels to be uncontaminated, we can plot the fluxes of an annulus in order from dimmest to brightest. The dimmer portion of pixels are expected to follow a characteristic curve that is produced from the Poisson distribution, while the brighter pixels will be seen to be significantly above this expected curve. Thus, fitting an ordered Poisson ensemble to the dim pixels provides an estimate of the GI flux. Note that this method does not consider the spatial ordering of the pixels, and is therefore insensitive to spatial correlations of the brightest Poisson-consistent dim pixels, which would be an indication of further non-GI contamination.

It would appear, then, that measuring the GI flux should require some combination of these two methods. However, we find that both methods give very similar estimates of the GI profile, so as to be essentially indistinguishable at the present level of development. This result supports the notion that non-Poissonities of pixels in an annulus are predominantly due to spatially-correlated hot structures. Thus, for this work, we present preliminary results for both methods and use the GI flux measured by the BDS test to demonstrate the consequences of these techniques for GI source modeling and dark matter constraints.

We now provide more detail about each method and present their estimates of the GI flux.

\subsection{Spatially Uncorrelated Annuli: The BDS Statistic}
\label{ssec:bds}
The BDS test \citep{Brock:1987,Brock:1991,Cromwell:1994,Brock:1996} was developed to test models of stochastic time series data by determining whether random residuals are indeed uncorrelated. Rejection of the null hypothesis (that model residuals are IID) would disfavor the model.

The test is as follows. Let $N$ be the number of pixels being tested. It is important that the pixels be ordered. For now, we order them as they are spatially arranged in the annulus, but we revisit this convention later in this section. Since the pixels form a circle, it does not matter which is chosen as first--it will still neighbor the last pixel in the sequence.

Denote $\{\phi_0,\phi_1,\ldots,\phi_{N-1}\}$ as the $N$ pixel fluxes. To test for the presence of spatial correlations over $m$ pixels, define $m$-dimensional flux vectors (there are $N$ of them), denoted $\{\Phi^m_0,\Phi^m_1,\ldots,\Phi^m_{N-1}\}$. For example, $\Phi^3_0=\{\phi_0,\phi_1,\phi_2\}$ and $\Phi^3_{N-2}=\{\phi_{N-2},\phi_{N-1},\phi_0\}$. 

This flux vector space with embedding dimension $m$ is promoted to a metric space by imposing the max norm. A threshold distance $\varepsilon$ is further specified to define the test function
\begin{equation}
	I_\varepsilon(\Phi^m_i,\Phi^m_j)=
	\begin{cases} 
	  1\quad\text{if $|\Phi^m_i-\Phi^m_j|\leq\varepsilon$,}\\
	  0\quad\text{otherwise.}
	\end{cases}
\end{equation}
A pair of vectors is considered ``correlated'' if all of their common components are within $\varepsilon$ of each other. The correlation integral is simply the average test function over all vector pairs
\begin{equation}
  C(\varepsilon, m, N)\equiv\frac{1}{N(N-1)}\sum_{i}\sum_{j\neq i}I_\varepsilon(\Phi^m_i,\Phi^m_j).
\end{equation}

Now, under the null hypothesis where the underlying probability distributions of the $N$ pixel fluxes are independent and identically distributed (IID), the ensemble average of the correlation integral satisfies
\begin{equation}
  \mean{C(\varepsilon, m, N)}=\mean{C(\varepsilon, 1, N)}^m.
\end{equation}
This follows from the fact that the test function over vectors is a product of scalar test functions
\begin{align*}
  \mean{I_\varepsilon(\Phi^m_i,\Phi^m_j)}& =\mean{\prod_{k=0}^{m-1}I_\varepsilon(\phi_{(i+k)\text{ mod }m},\phi_{(j+k)\text{ mod }m})} \\
  & =\prod_{k=0}^{m-1}\mean{I_\varepsilon(\phi_{(i+k)\text{ mod }m},\phi_{(j+k)\text{ mod }m})} \\
  & =\mean{I_\varepsilon(\phi_i,\phi_j)}^m
\end{align*}
where the second line follows from the independence of the pixels, and the third line follows from each pixel sharing the same underlying probability distribution for the observed flux.

For IID pixels, the BDS statistic
\begin{equation}
  w(\varepsilon, m, N)=\sqrt{N}\frac{C(\varepsilon,m,N)-C^m(\varepsilon,m,N)}{\sigma(\varepsilon,m,N)}
\end{equation}
is proven to be asymptotically standard normal distributed. Accurate fitting functions for the statistical width $\sigma$ of $C(\varepsilon,m,N)$ were determined with simulations, and are reproduced in \app{app:BDSsigma}.

Efficient open-source algorithms for carrying out BDS tests are available \citep{LeBaron:1997,Franch:2002}. We use a version of the LeBaron C code\footnote{The source code is available with the supplementary material in the online version of \citet{LeBaron:1997}.}, modified to analyze a circular series of data.

The BDS test has three parameters: the threshold distance $\varepsilon$, the embedding dimension $m$, and the statistical threshold $N_\sigma$ at which the null hypothesis of IID pixels is rejected. In addition, we need to identify a threshold pixel count $N_{\text{th}}$ below which the BDS test becomes unreliable due to a lack of sufficient statistics.

\citet{Brock:1991} finds that $w$ has an approximately standard normal distribution for $N_{\text{th}}=500$, $m\leq5$, and $\sigma_\phi/2\leq\varepsilon\leq2\sigma_\phi$, where $\sigma_\phi(\psi)$ is that standard deviation of the pixel fluxes at GI radius $\psi$. Simulations of astrophysical distributions by \citet{Baxter:2011rc} show that for astrophysical distributions and the level of precision required for our purposes, $N_{\text{th}}=50$ is sufficient to produce accurate results. We use $\varepsilon=\sigma_\phi/2$, and carry out 4 BDS tests with $m$ taking values 2, 3, 4, and 5. The null hypothesis is rejected if |w|>3, corresponding to a greater than $3\sigma$ deviation from IID.

If the null hypothesis is rejected for a given set of pixels, then the brightest pixel is removed/masked, and the test is repeated with one fewer pixel. Thus, the bright contaminated pixels are removed until the remaining data are consistent with being IID to within $3\sigma$, according to BDS. The BDS flux is then determined to be the median flux of the remaining pixels.

If during this process the number of remaining pixels drops below $N_\text{th}=50$, then the BDS flux is taken to be undetermined. Likewise, the BDS flux is undeterminable for annuli with fewer than $N_\text{th}$ pixels.

\begin{figure*}
  \includegraphics[width=0.5\linewidth]{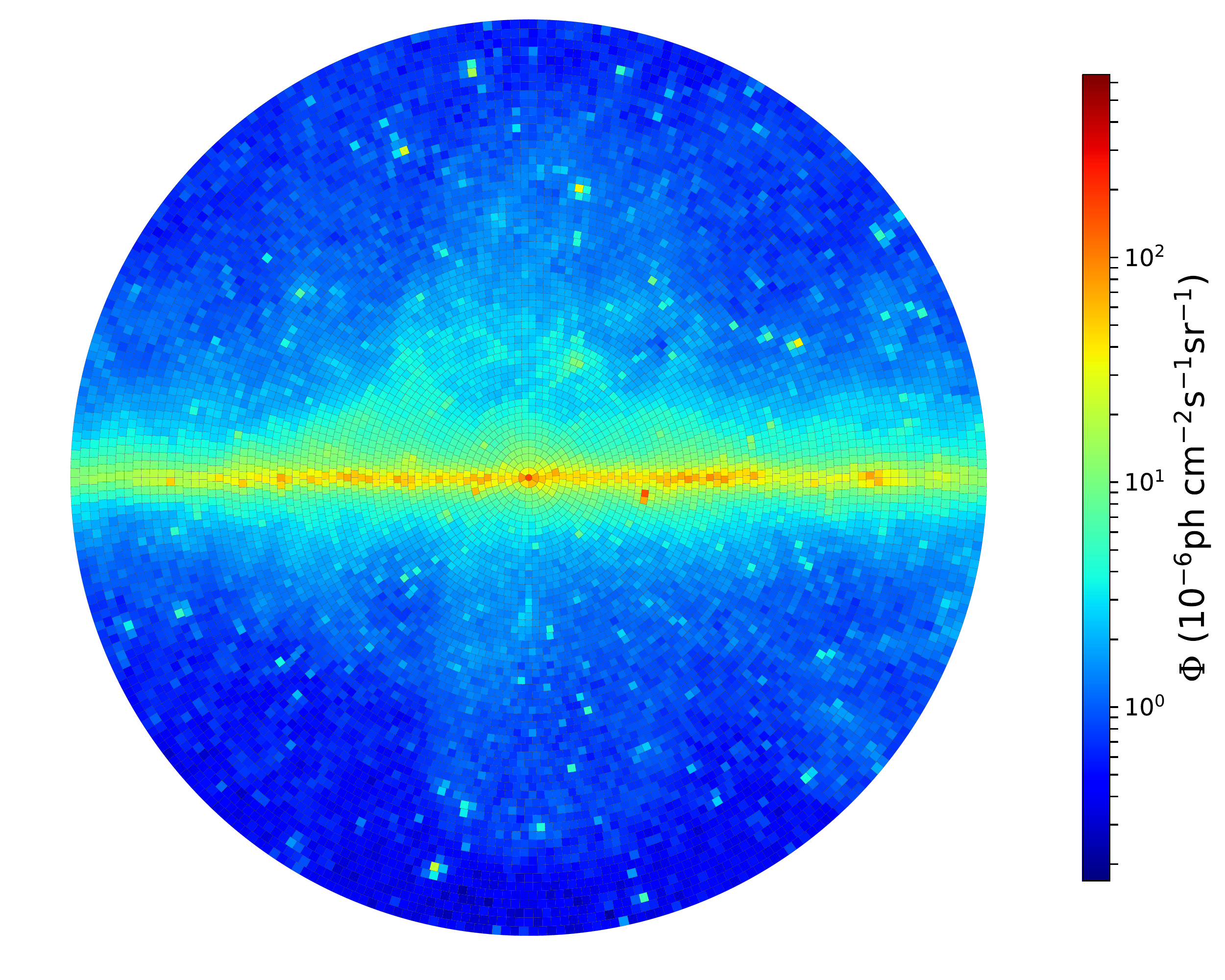}\includegraphics[width=0.5\linewidth]{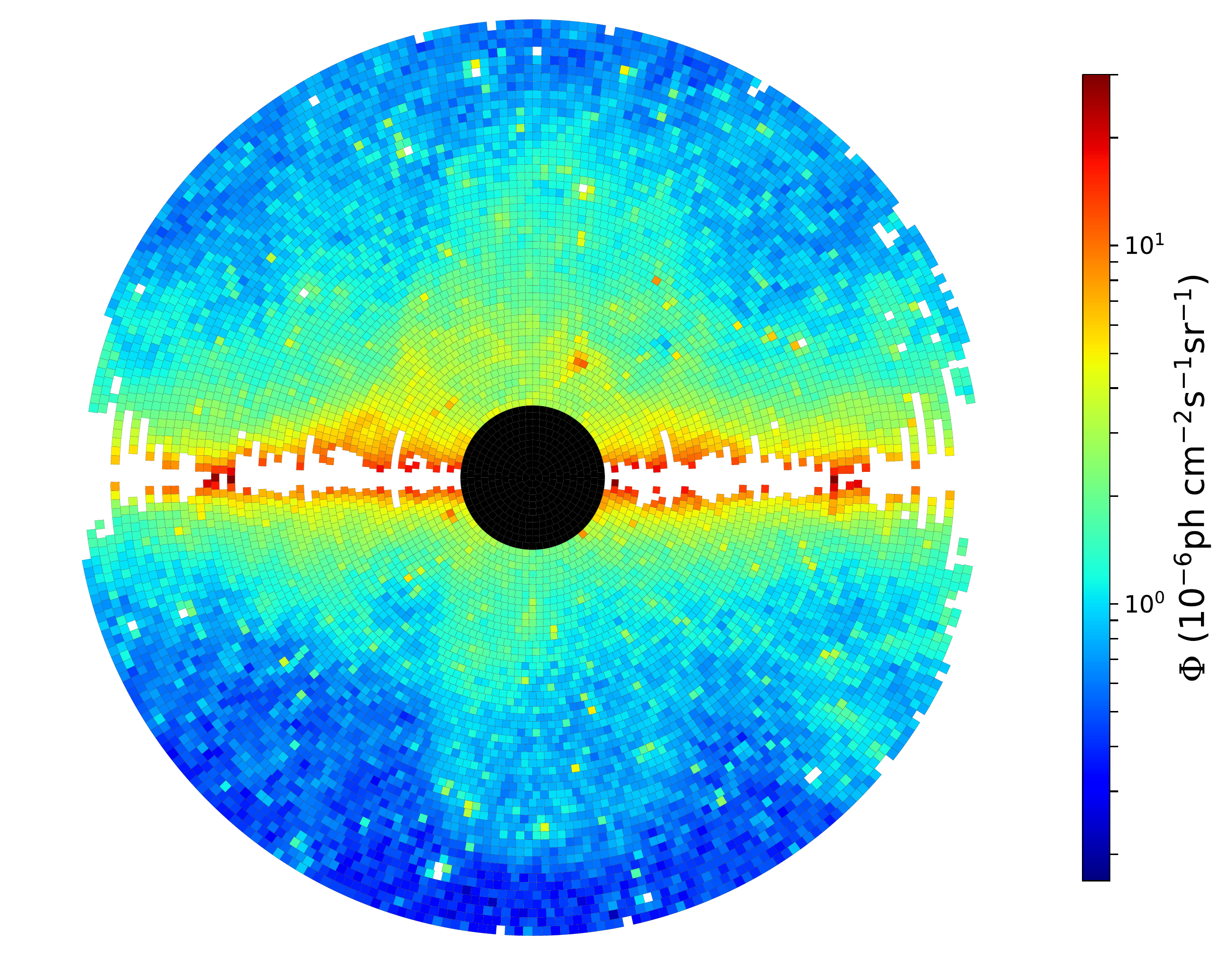}
  \caption{\label{fig:GItiling1}Left: GI tiling of the $\gamma$-ray sky at energies 2--300 GeV at $1^\circ$ resolution out to $60^\circ$. Right: Same except with white regions indicating pixels that were removed by the BDS analysis. Notice that, with this procedure, significant residual structure remains unremoved. Black annuli indicate failed BDS tests, in this case from too few pixels.}
\end{figure*}

In the right panel of \fig{fig:GItiling1}, we show the pixels that are masked by the BDS test for the $1^\circ$ GI tiling. Annuli that are completely masked have an undetermined BDS flux. While much of the Galactic disk, bright point sources, and extended gas cloud features are removed, we can still see that much non-GI structure remains. Some annuli where the disk feature is a bit thicker and also contains bright point sources appears to make the bright pixels numerous enough to be reasonably consistent with being upward fluctuations of an IID-distributed ensemble of pixels. Fewer bright, contaminated pixels are removed from such annuli.

At smaller pixel size, we find that annuli with larger GI radii will have undetermined or zero BDS flux. This is attributed to a large number of empty pixels at high Galactic latitudes where fluxes are small. As more bright pixels become removed, the relative positions of empty pixels become more likely to be correlated (as with the pixels with only 1 count, or 2 counts, etc.). The quantization of flux runs counter to the assumption that the pixels are sampled by IID continuous distributions. Therefore, the failure of the BDS test at high GI radii with small pixels can be attributed to the inapplicability of the BDS test in this regime, where it has a tendency to identify anomalous spatial structure in regions of low flux.

\begin{figure}
	\center
  \includegraphics[width=0.5\linewidth]{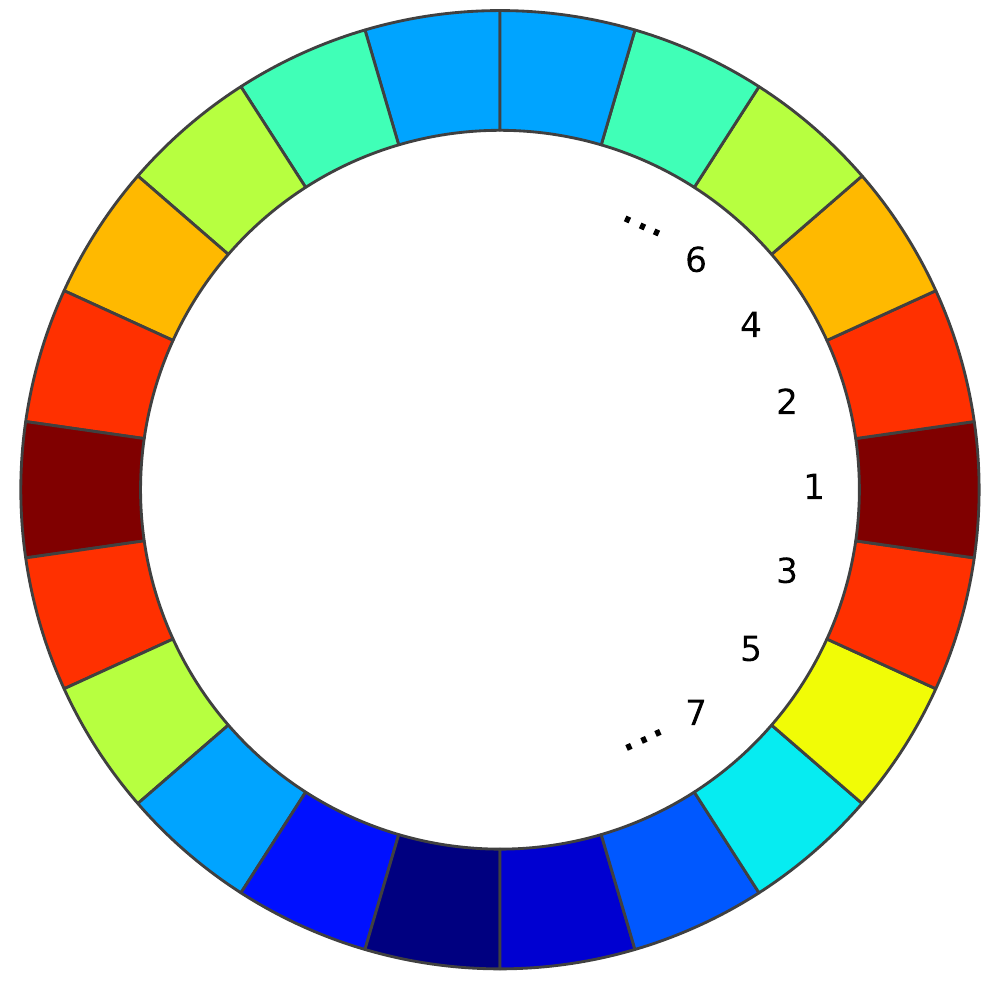}
  \caption{\label{fig:interleave}Schematic of the pixel interleaving procedure. Shown here are pixels of an annulus. The pixels are re-ordered as indicated by numeric labels, beginning at the Galactic plane (labelled 1), and alternating between the next North pixel and next South pixel. Due to the North being consistently brighter, interleaving results in alternating bright-dim-bright-dim patterns until the brighter pixels are removed.}
\end{figure}

We introduce an innovation that allows the BDS test to remove much more of the observed structure, while simultaneously improving its stability at high radii. This is achieved by applying a North-South interleaving of the pixels, as demonstrated in \fig{fig:interleave}. Since the GI component we are attempting to extract is uniform, it is unaffected by the ordering of the pixels. However, observed spatial correlations are affected by the re-ordering the pixels. In our sky, there tends to be more extended structure from gas clouds in the Northern hemisphere. Thus, when interleaved with relatively clean pixels in the Southern hemisphere, alternating bright-dim pixel patterns are created which persist until all of the offending bright pixels are removed. Not only does this make the removal of structure much more effective, but it is also observed to improve the stability of the BDS test in the presence of many empty and low-count pixels.

\begin{figure}
  \includegraphics[width=\linewidth]{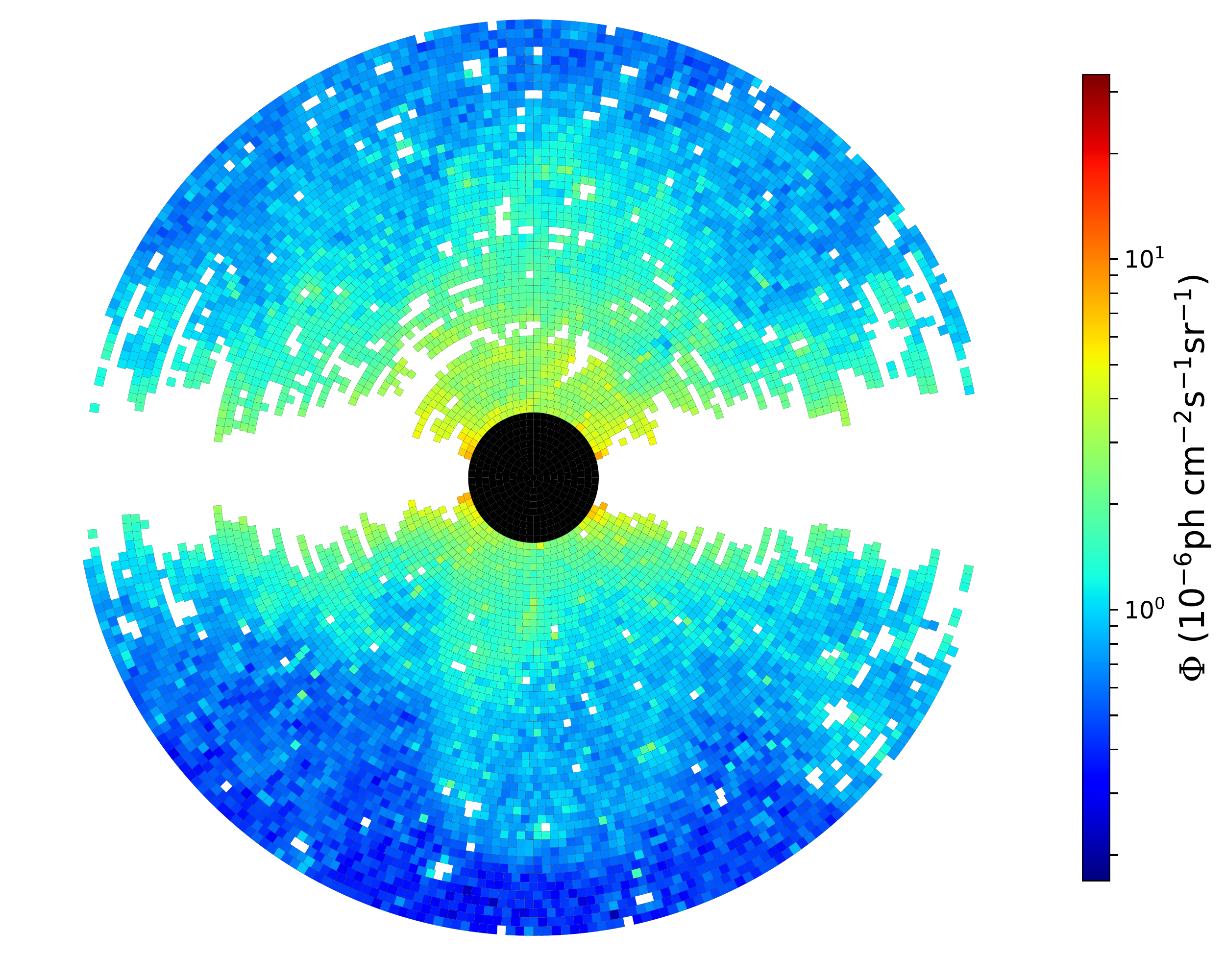}\\
  \caption{\label{fig:bdspixil}Results of the BDS tests after interleaving pixels on the inner $60^\circ$ with $1^\circ$ pixels.}
\end{figure}

We can see in \fig{fig:bdspixil} how much additional structure is removed when applying the BDS test to interleaved pixels. The magnitude of the BDS flux is shown by the lower lines in \fig{fig:fluxplot} for our three pixel sizes. The $1^\circ$ pixels are large enough to not contain many low-count/empty pixels, and so the BDS test remains stable throughout the large radii. Spikes caused by bright sources are completely removed, and the remaining flux is reduced by a factor of 2--3, except within $\sim 10^\circ$ of the Galactic center, and $\sim 30^\circ$ fo the anti-Galactic center where the flux is not reduced as efficiently. In those regions, the small pixel GI tilings are very stable and extend the measurement of GI flux to smaller radii from the Galactic and anti-Galactic centers.

\begin{figure}
  \hspace*{-10pt}\includegraphics[width=1.15\linewidth]{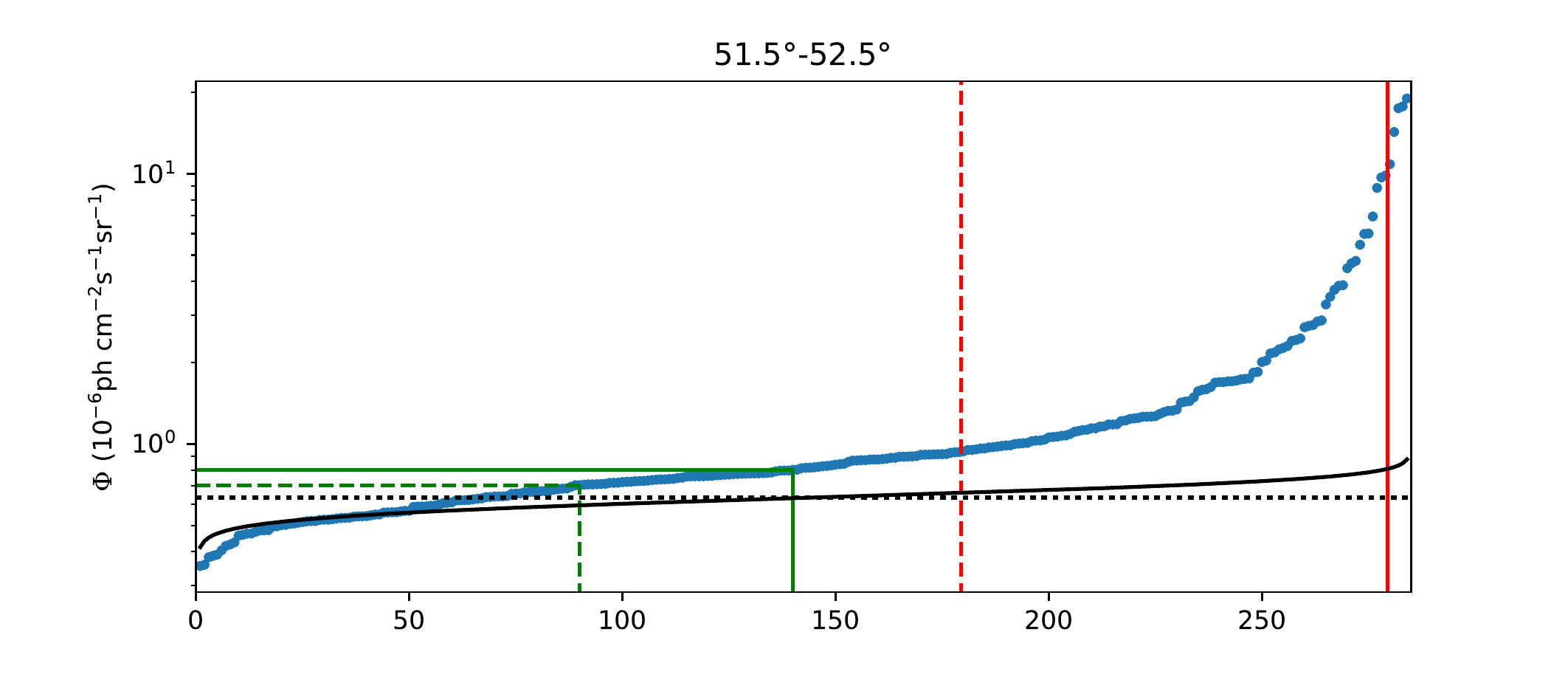}\\
  \hspace*{-10pt}\includegraphics[width=1.15\linewidth]{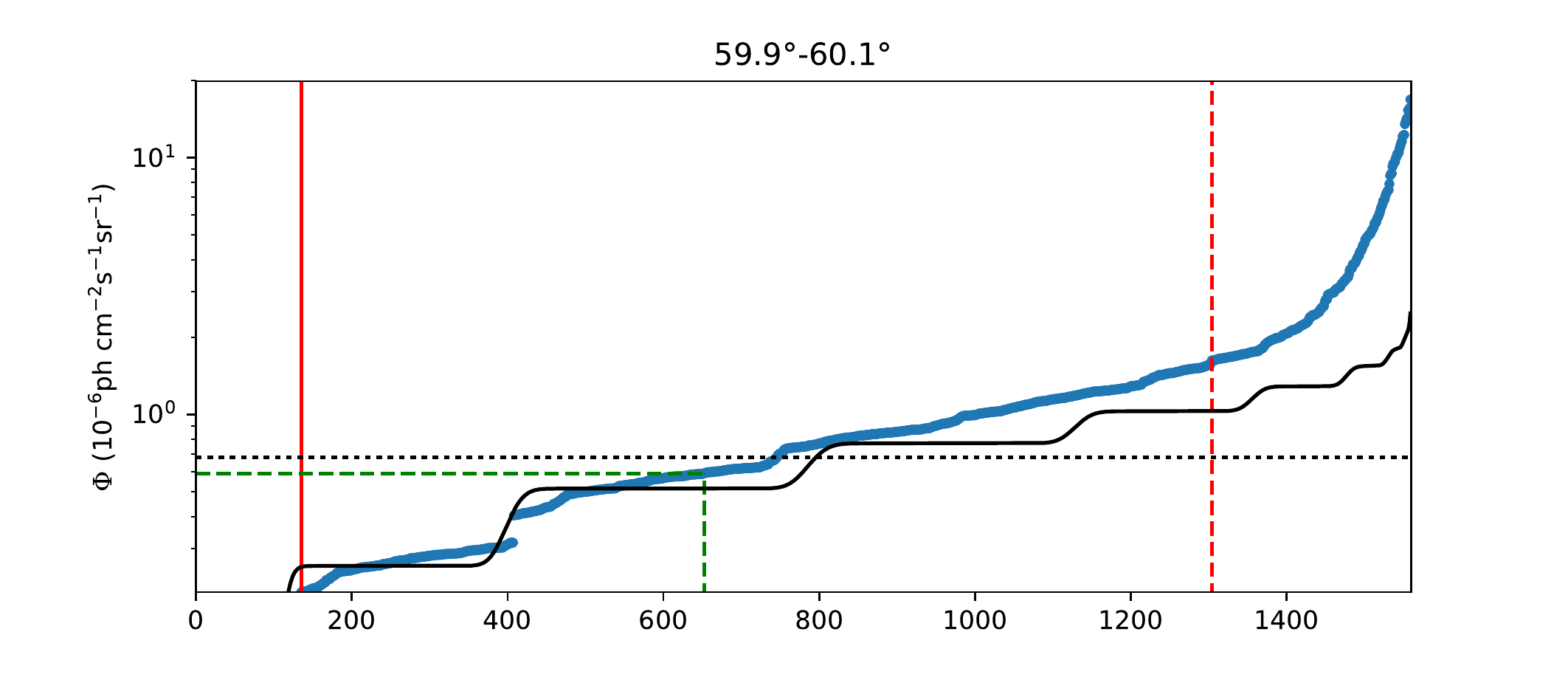}\\
  \hspace*{-10pt}\includegraphics[width=1.15\linewidth]{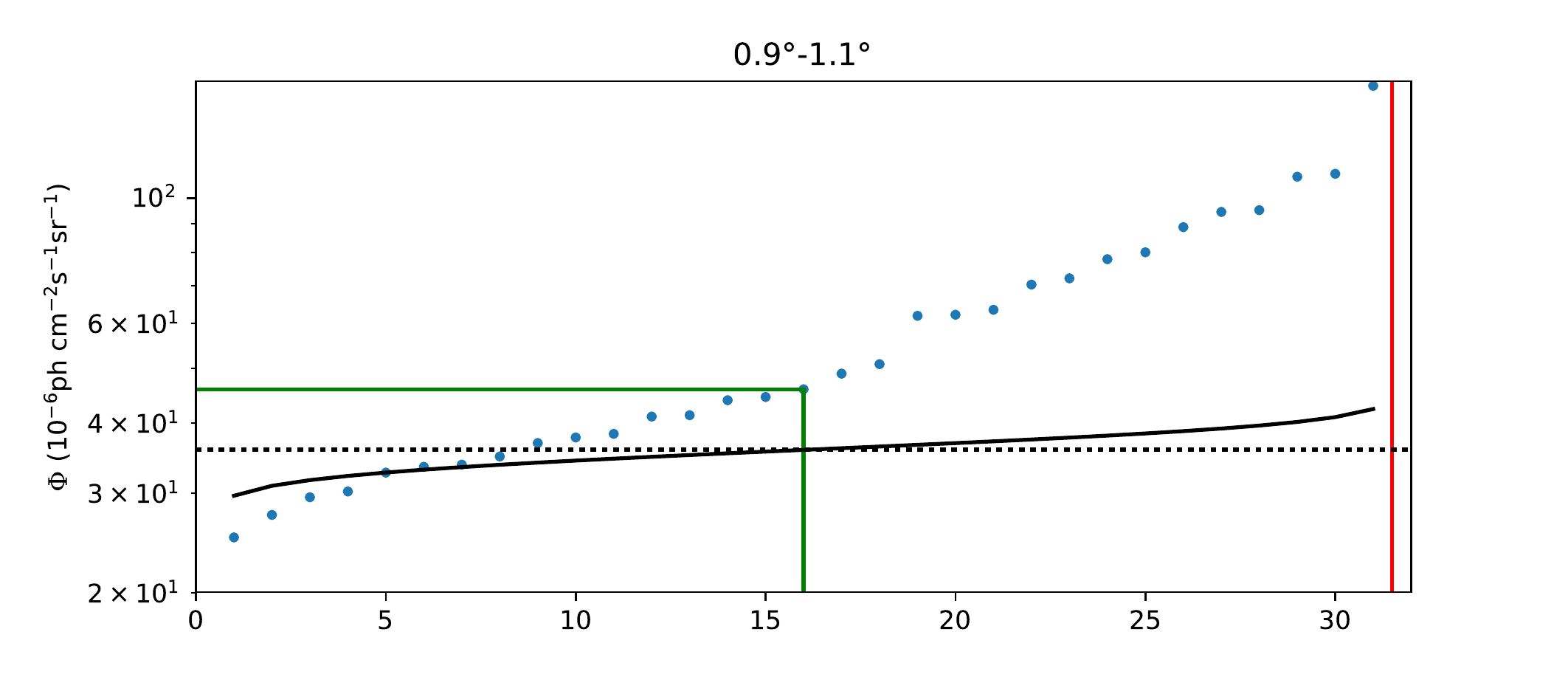}
  \caption{\label{fig:ordered}Ordered pixel fluxes for three representative annuli (blue dots). The bright pixels to the right of the red vertical lines are the ones cut by the BDS test (dashed line for interleaved pixels). The green lines indicate the median remaining pixel and corresponding flux. The black solid line shows the mean ordered Poisson ensemble for the flux indicated by the black dotted line. Top: The $52^\circ$ annulus with $1^\circ$ pixels. Note how the BDS test with interleaving pixels removes many more pixels, but barely affects the GI flux estimate. The Poisson ensemble fit produces nearly the same estimate of the GI flux. Middle: The $60^\circ$ annulus with $0\fdg2$ pixels. The dimmest $\sim150$ pixels are empty and the next $\sim250$ pixels contain one photon. Interleaving pixels stabilizes the BDS test for this annulus. The Poisson fit clearly shows the quantization of the photon occupation of the pixels, and its fit is near the GI flux estimate of the BDS test. Bottom: The Poisson fit allows for GI flux estimates near the Galactic center where annuli have few pixels. This $1^\circ$ annulus with $0\fdg2$ pixels cannot estimate a GI flux using the BDS test.}
\end{figure}

One might have expected that the removal of additional structure would result in a significantly smaller median pixel flux, and thus a dimmer measured BDS flux. However, we see that the reduction in the BDS flux is marginal. The reason for this is straightforward to understand when we consider the ordered pixel ensemble as shown in \fig{fig:ordered} (top) for annulus at GI radius $\psi=52^\circ$ for the $1^\circ$ GI tiling.

When $N$ samples of a Poisson distribution are taken (assuming the expected count parameter is significantly greater than 1), most of those samples will have values clustered around the peak of the distribution, while there are expected to only be a few with significantly smaller values, and a few with significantly larger values. This consistently generates the characteristic horizontal ``S''-shaped curve shown by the solid black line in the top example of \fig{fig:ordered}. Contamination from bright sources skew the bright end of the curve to be significantly brighter than the Poisson expectation. Assuming that the contaminated bright pixels do not represent more than half of the pixels, then the median of the remaining pixels still remains very close to the peak of the Poisson distribution that agrees with the dim pixels.

In fact, if the dim half of the pixels is indeed consistent with being the dim half of an ordered ensemble of IID Poisson pixels, then the median pixel is a good estimator of the underlying Poisson flux. The BDS flux is slightly lower by taking the median pixel after removing structure, but this change remains marginal. However, if more than half of the pixels are found to be significantly skewed from the Poisson expectation, then the median pixel would be contaminated and would not provide a sufficient estimate of the GI flux--removal of structure with the BDS test would be a necessary procedure to determining the GI flux.

This consideration leads to an alternative to the BDS test for estimating the GI flux. Assuming that the dimmest pixels are IID samplings of a Poisson distribution, we can fit them with a typical Poisson ordered-ensemble and determine a best-fit flux to the underlying uniform Poisson distribution. While this method no longer considers spatial correlations (since ordering the pixels by flux erases their spatial ordering), it does enforce that the underlying uniform probability distribution of the uncontaminated pixels be consistent with Poisson. The estimation of the maximal uniform Poisson flux of an annulus is now described.

\subsection{Uniform Poisson Annuli: Ordered Ensemble Statistics}
\label{ssec:poisson}

This second method for determining the GI component of $\gamma$ rays is intended to be applied in the following scenario. For a given annulus of $N$ pixels, imagine that every pixel contains a common flux component (the GI-component), and that some of the pixels contain additional fluxes which we will refer to as contaminations.

Now the GI flux received in each of the $N$ pixels will be independent and random, sampled from a common Poisson distribution. When a $N$ randomly sampled Poisson variables are ordered, they closely follow characteristic curves associated with the underlying probability distribution, as shown by the solid black lines in \fig{fig:ordered}. Analytic formulas for these average ordered ensembles are derived in \app{app:aveordered}. Since the exposure map is close to uniform within each annulus, we calculate ordered ensemble statistics under the assumption of constant exposure (using the average exposure of the annulus).

When the pixel fluxes are such that the counts per pixel are much greater than 1, then we generate the characteristic horizontal ``S'' curve, as in the top panel of \fig{fig:ordered}. In contrast, low flux ensembles with empty and low-count pixels form a series of quantized steps as in the middle panel of \fig{fig:ordered}. The fact that the data in each step of that figure is not constant is indicative of the exposure map's slight variation. We can see that the exposure variations do not affect the estimation of the GI flux--it is the positions of the transitions to the next count that are important for estimating the underlying flux. Unlike the BDS test, the ordered ensemble method can confidently be used on annuli with only a few pixels as seen in the bottom panel of \fig{fig:ordered}.

We see that the fluxes of the ordered ensemble curves (indicated by the horizontal, dashed, black lines) provide ordered pixel fluxes that are good fits to the dim pixels, but underestimate the brightest pixels. This is consistent with the dim pixels being determined by an underlying GI flux, and with the brighter pixels being contaminated by bright, localized sources.

The shown fits to the GI fluxes are conservative for two reasons. First, they are ensured to overestimate most of the dimmer data. Secondly, they ignore the fact that some of the contaminated pixels had GI contributions that were downward statistical fluctuations (dimmer than average), but were removed from their position in the ordered pixels and pushed to the right to the contaminated pixels. Accounting for this would cause the rise of the ordered ensemble curve to be slower, and the estimated GI flux would be smaller.

One robust way to estimate the GI flux in this situation is to determine a likelihood function for the ordered Poisson ensemble. Then a best-fit GI flux and conservative 95\% upper bound flux could be precisely determined from the data.

Once pixel counts are ordered, the counts are no longer independent. If one pixel deviates randomly (for example, the fifth dimmest pixel is brighter than is average for being fifth dimmest), then the neighboring ordered pixels are more likely to deviate in the same direction. Conditional probability distributions for the counts of ordered pixels given the counts of dimmer pixels are derived in \app{app:condordered}. The conditional probabilities are necessary to define the likelihood function for the ordered ensemble.

The likelihood function is presented in \app{app:likelihood} as a function of the GI flux. For an annulus of $N$ pixels, it determines the likelihood that the first $N_{\text{th}}$ pixel counts could have been produced from a given GI flux. $N_{\text{th}}$ is a threshold count used to determine how many pixels are consistent with being uncontaminated. If contaminated pixels are included, the fit becomes worse and the likelihood goes down compared to fits with only uncontaminated pixels. Such likelihood functions are necessary for automatic GI-flux fitting routines. The development of such subroutines will be carried out in future work.

\begin{figure}
  \includegraphics[width=\linewidth]{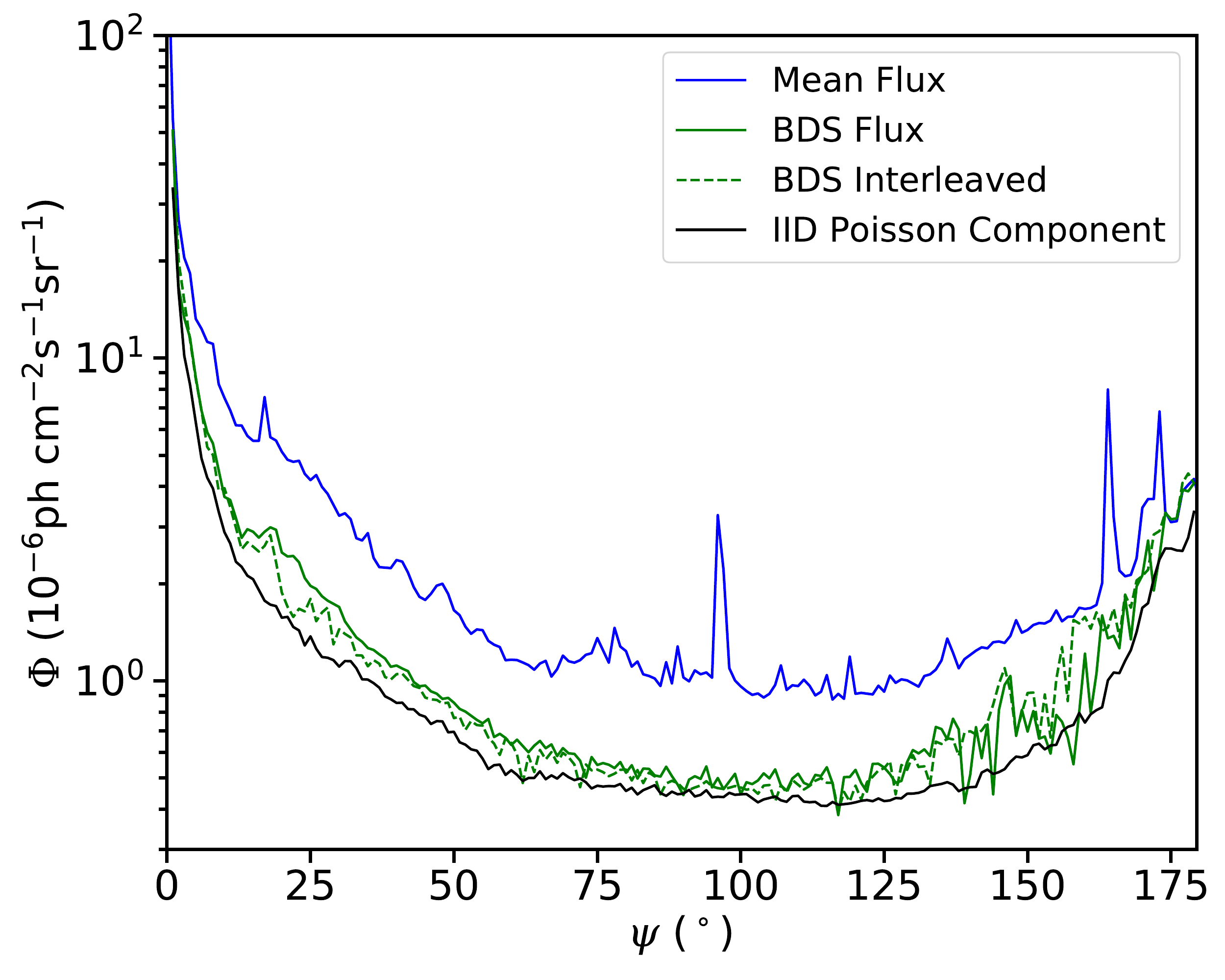}\\
  \includegraphics[width=\linewidth]{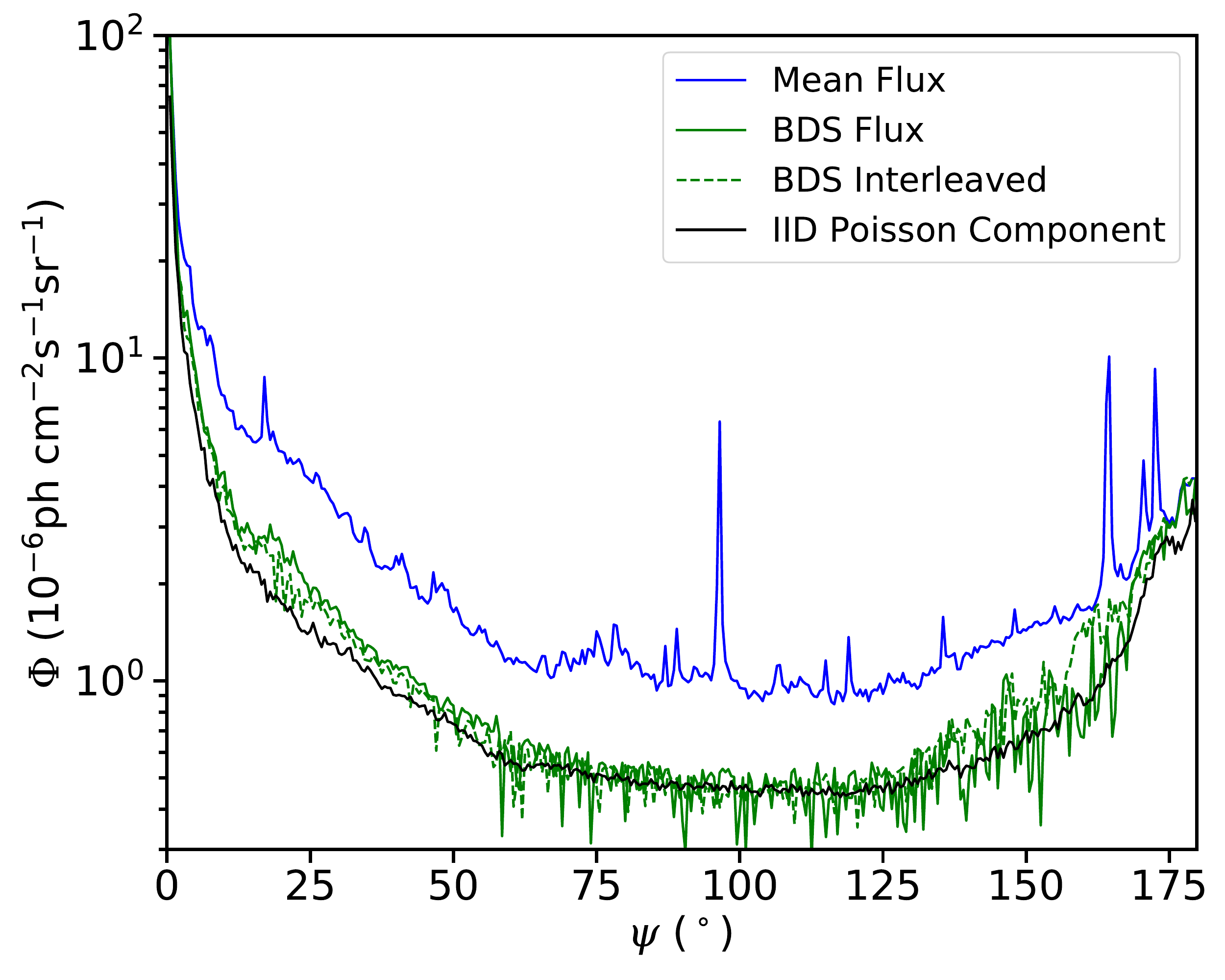}
  \includegraphics[width=\linewidth]{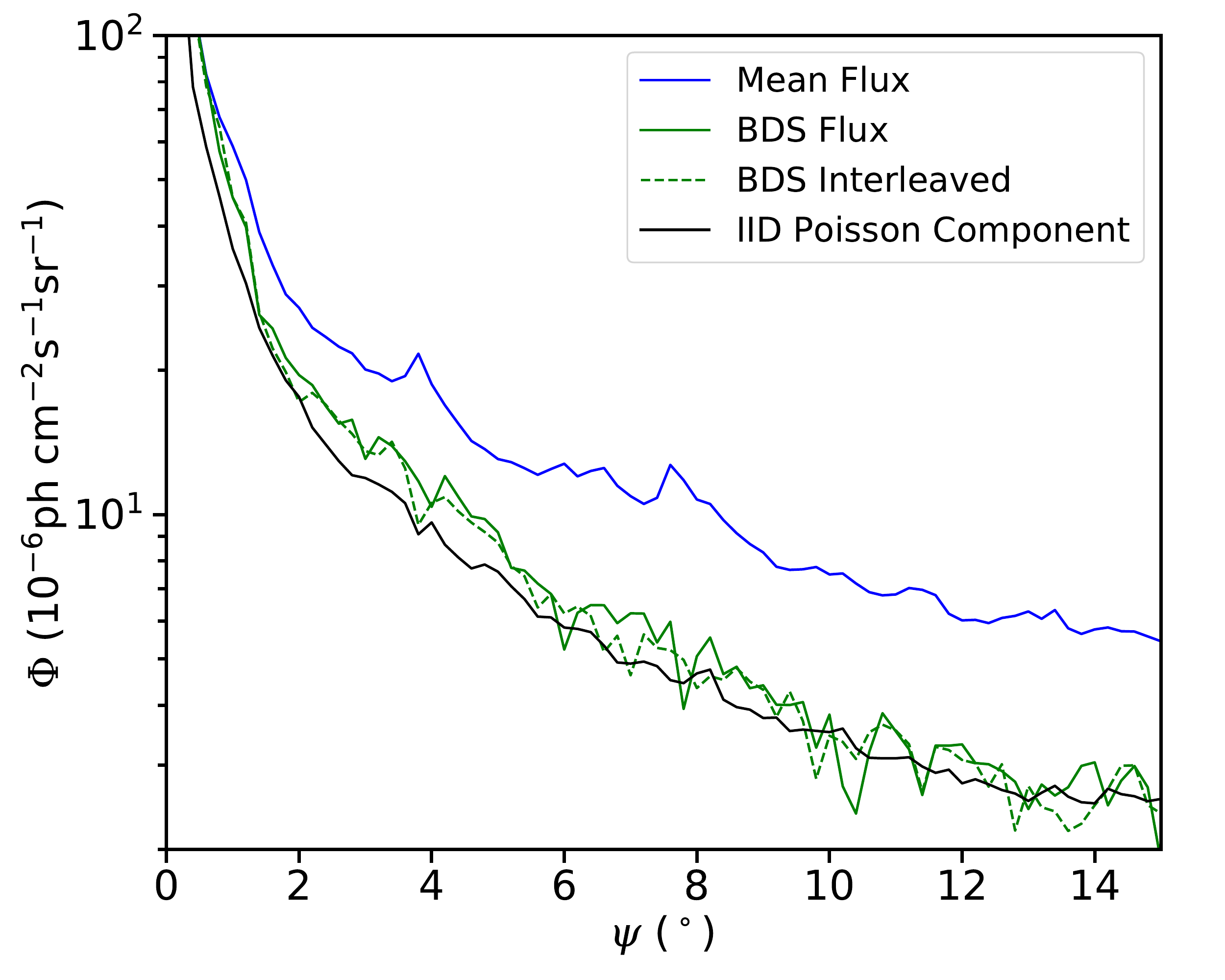}
  \caption{\label{fig:poissonflux} Comparison of the GI flux determined with the different methods at each resolution. The mean flux profile with $1^\circ$ resolution is also shown in each. Top: $1^\circ$ resolution. Middle: $0\fdg5$ resolution. Bottom: $0\fdg2$ resolution over the inner $15^\circ$. This shows the potential reach to estimate the GI profile as close as $1^\circ$ from the Galactic center. Beyond $15^\circ$, the BDS tests are unstable, and the ordered Poisson ensemble flux is uncertain until proper statistical fitting procedures are developed.}
\end{figure}

For the present study, we conservatively fit the dim pixels with average ordered Poisson ensembles by hand. The resulting GI-flux profile is shown in \fig{fig:poissonflux}, compared with the GI flux found with the BDS test. We see that the two techniques produce very similar results in regions where the BDS test is stable. This suggests that the spatially correlated pixels detected with the BDS test are indeed the contaminating non-Poissonities seen by the ordered ensemble. 

Two ways that can account for some of the observed differences are as follows. First, the BDS test does not test that the underlying IID distribution is Poisson. Thus, IID non-Poissonities could persist in the GI flux measurement using the BDS test, which would cause the GI componenet to be overestimated. For the second case, the ordered ensemble does not retain spatial information of the pixels, and so spatial correlations of the Poisson-consistent pixels could be present which would indicate the presence of dim localized structure which would also cause a slight overestimation of the GI flux.

Where the BDS flux is stable, the ordered Poisson method usually extracts a lower GI flux, as much as about $30\%$, and its profile is less noisy than the BDS profiles. The stability of the ordered Poisson method suggests that any observed dips of more the $30\%$ in the BDS flux at a single annulus can confidently be identified as instabilities rather than real features in the GI flux profile.

In general, the most stable of the two methods is the ordered ensemble test. It is able to be used confidently with annuli having few pixels, or having many low-count pixels. However, a precise, statistical estimation of the GI flux using this method requires further development. In any case, approximate results are consistent with the results of the BDS test in the annuli where BDS is stable.

Thus, for the remainder of this paper, we will use the BDS test results to estimate and model the GI flux. Based on the observations of this section, the BDS-derived GI flux of the observed $\gamma$ rays is taken to be a conservative overestimate of the GI flux, and any instability dips that affect the analysis will be adjusted.

\section{Source Modeling and GI Component}
\label{sec:model}

\subsection{Spatial Templates and Mock Data}
\label{ssec:mocks}

The diffuse galactic background is composed of gamma-ray emission from inverse Compton scattering, bremsstrahlung, and $\pi^0$ decays. 
We used the WebRun interface of the GALPROP cosmic ray propagation code~\citep{Porter:2008ve, Vladimirov:2010aq} to model the diffuse gamma-ray emission from these three galactic components.
The predicted spatial distribution and spectra for these components are dependent on (1) the assumed distributions of cosmic ray sources and gas as well as (2) the diffusion and propagation parameters input into the cosmic ray propagation code.
We assumed the set of parameters as given in `model A' of \citet{Calore:2014xka}. This model was chosen because it was shown by the authors of \citet{Calore:2014xka} to be self-consistent in the sense that the component spatial templates' best-fitting spectra when fit to the data were in good agreement with the spectra predicted by GALPROP.

The input parameters from model A in \citet{Calore:2014xka} are as follows. For a full description of each parameter, see \citet{FermiLAT:2012aa} and \citet{Carlson:2016iis}. The cosmic ray diffusion is characterized by scale radius $r_D=20$ kpc, scale height $z_D=$4 kpc, Alfv\'{e}n speed $v_A=32.7$ km s$^{-1}$, diffusion coefficient $D_0=5.0\times10^{28}$ cm$^3$ s$^{-1}$, convection velocity gradient $dv/dz=50$ km/s, electron (proton) injection spectrum power law index $\alpha_e(\alpha_p)=2.43 (2.47)$, electron (proton) injection spectrum normalization $N_e (N_p)=2.0 (5.8)$ at 34.5 (100) GeV, gas spin temperature $T_S=150$ K, and optical/IR/CMB radiation field normalizations 1.36/1.36/1.0. The exponential magnetic field model is characterized by $B_0=9\ \mu$G, scale radius $r_c=5$ kpc, and scale height $z_c=2$ kpc.
The cosmic ray source distribution is based on the measured supernova remnant distribution~\citep{Case:1998qg}. (Note that this distribution falls with radius towards the galactic center, and likely underestimates the cosmic ray source density in the innerpost kiloparsec of the Milky Way~\citep{Carlson:2016iis}.)

The resulting models of inverse Compton/brems\-strah\-lung/$\pi^0$ emission were then input into the \textit{gtobssim} tool from the Fermi Science Tools package to simulate the contribution of the photon events from these galactic astrophysical backgrounds to the total observed gamma-ray data. These mock observations were generated using the same time range and spacecraft pointing as the actual dataset. The normalization of each background model component was chosen to match the best-fit normalization when this same set of GALPROP background model templates was fit to the data in \citet{Horiuchi:2016zwu}. This fitting is done in a $20^\circ\times20^\circ$ field-of-view about the Galactic center. The resulting mock observations are then binned into pixels and energy bins.

In addition, the emission from resolved point sources is modeled using the spectral parameters in the 3FGL catalog \citep{Acero:2015hja}. Rather than simulating mock observations of the point sources with gtobssim, we model each source's flux for each energy bin by integrating over the analytic fit to the spectrum. The model for the instrument point spread function is detailed in \app{app:psf}.

Finally, our model includes the contribution of the isotropic $\gamma$-ray background (IGRB). In \fig{fig:fluxplot}, the magnitude of the IGRB is indicated for the range of models of cosmic ray sources and propagation parameters explored in \citet{Ackermann:2014usa}. Analyses of this radiation indicate that it is dominated by populations of unresolved point sources \citep{Lee:2015fea,Fornasa:2016ohl}. This provides an important component of the GI profile at large GI radii. For our analysis, we use the best-fit model produced with the foreground model described above. The resulting IGRB model is provided in \app{app:igrb}. We note that it is brighter than the range of models in \fig{fig:fluxplot} from previous fits to the IGRB.

\begin{figure}
  \includegraphics[width=\columnwidth]{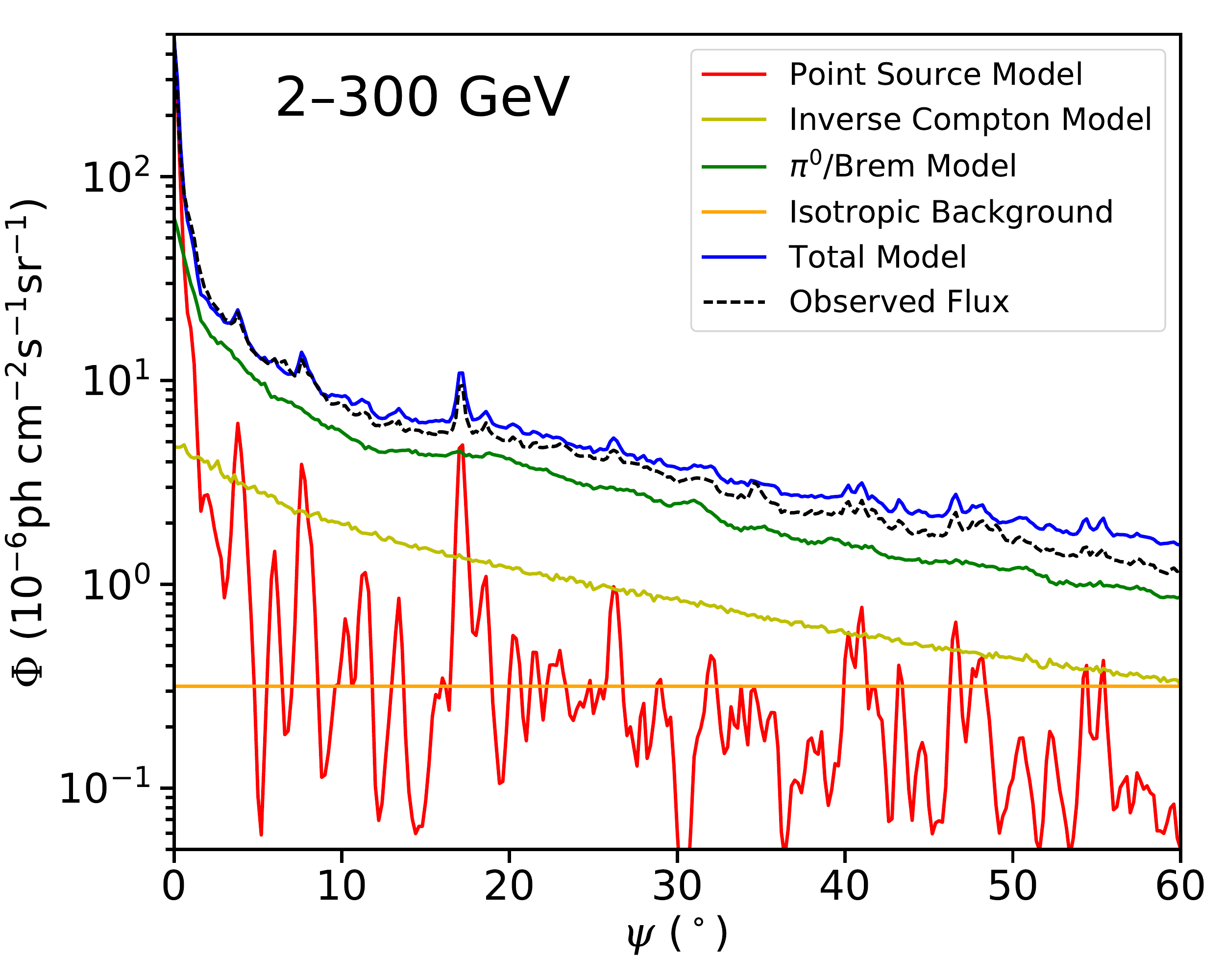}
  \caption{\label{fig:meanmodels} GI radial average flux profile of our $\gamma$-ray sky model, shown with the observed average flux profile for comparison. The sky model parameters were taken from Model A of \citet{Calore:2014xka}, with component normalizations determined in a $20^\circ\times20^\circ$ field-of-view centered on the Galactic center in \citet{Horiuchi:2016zwu}.}
\end{figure}

The mean flux in each annulus for the $0\fdg2$ GI tiling is shown for each model component in \fig{fig:meanmodels}. Also shown is the sum of the model components and the observed profile. We see that our model provides a good fit to the observed mean profile at radii less than about $9^\circ$, and overestimates the flux at larger radii out to at least $60^\circ$.

\begin{figure*}
	\includegraphics[width=\columnwidth]{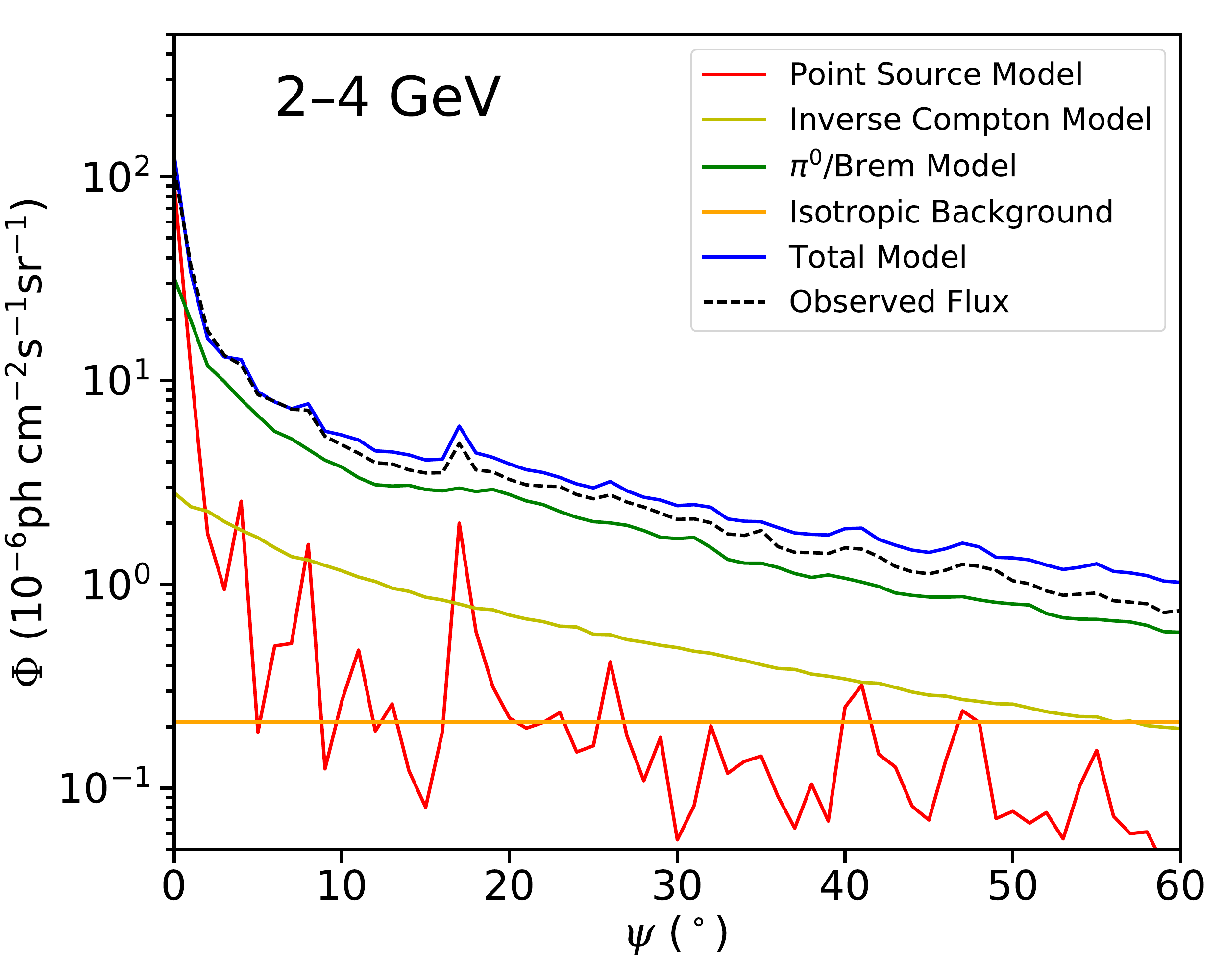}
	\includegraphics[width=\columnwidth]{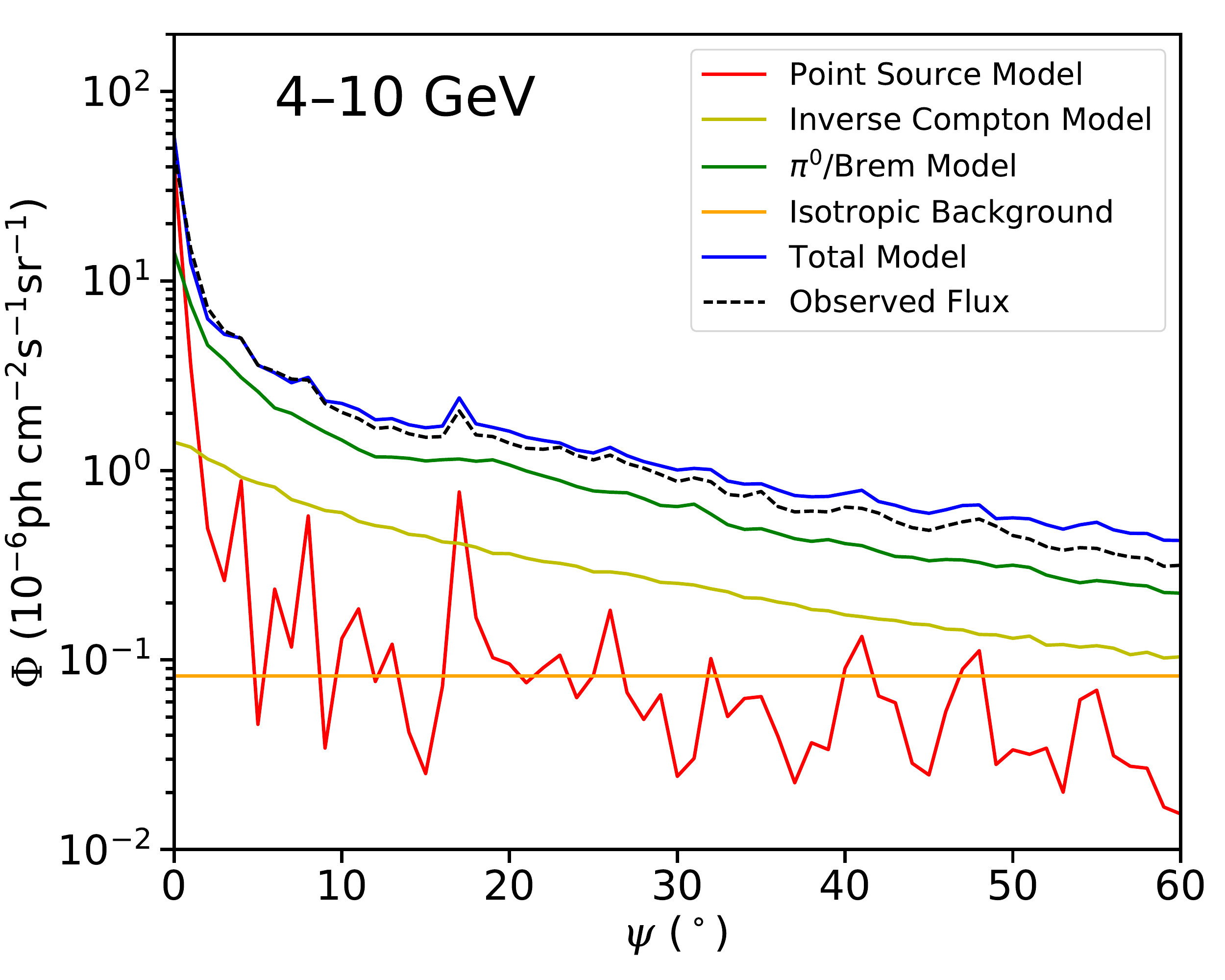} \\
	\includegraphics[width=\columnwidth]{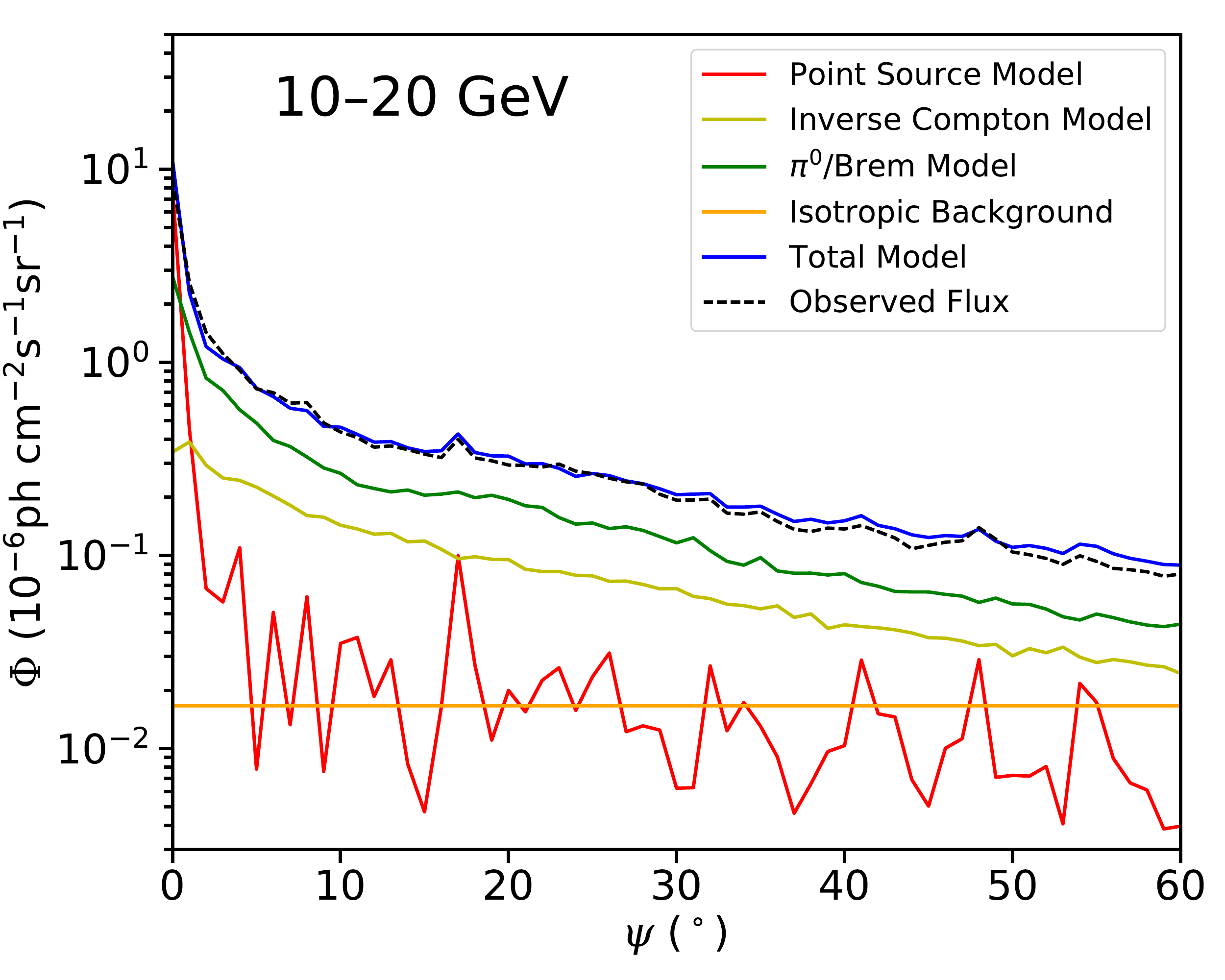}
	\includegraphics[width=\columnwidth]{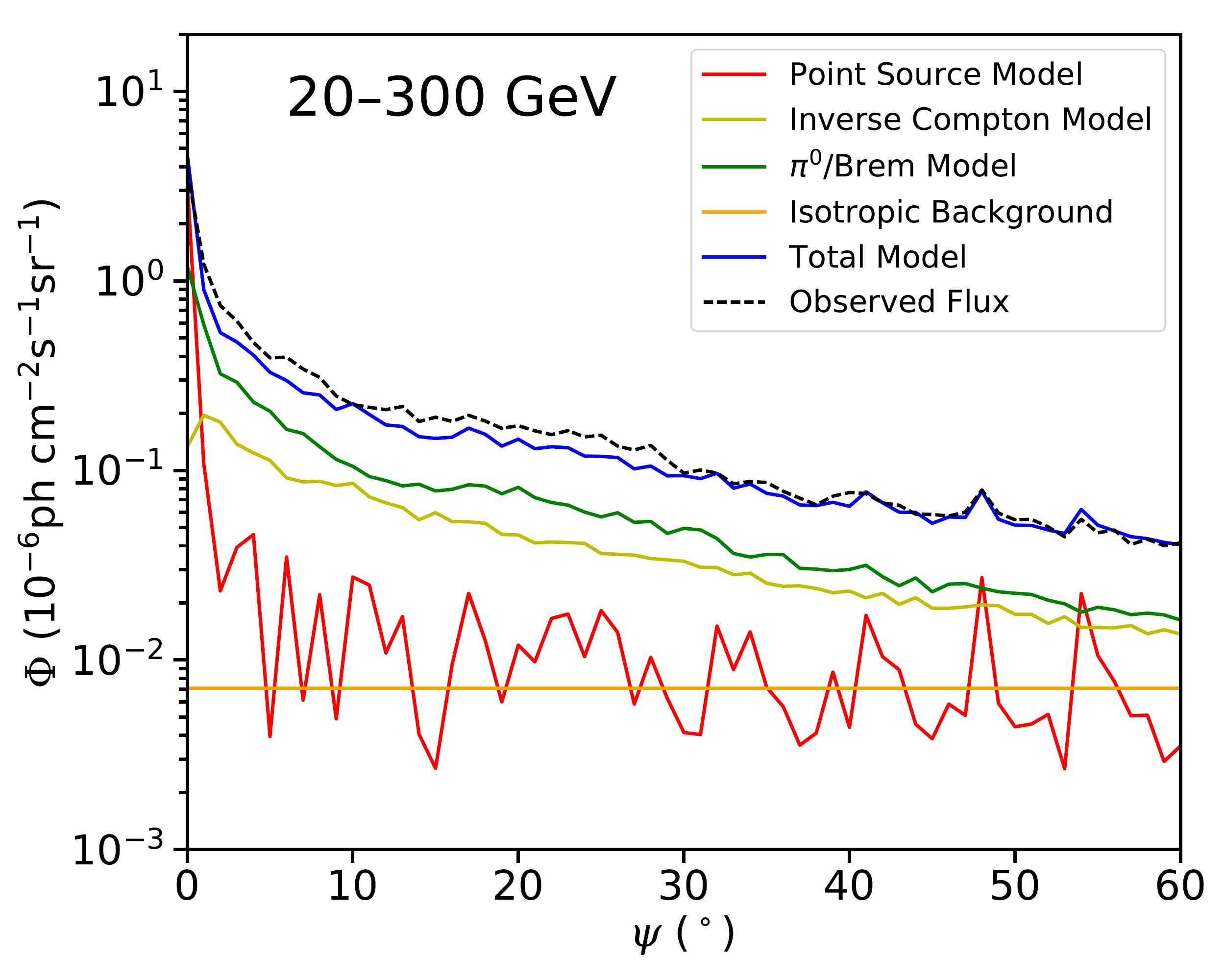}
	\caption{\label{fig:meanmodelbins} Mean flux profile for our model and observations in 4 energy bins. Note the disagreement between observations and model at low (high) energies is predominantly at high (low) radii.}
\end{figure*}

To ensure that our source model gives a consistent spectrum, we consider the model and data in four energy bins: 2--4, 4--10, 10--20, and 20--300 GeV. These are shown in \fig{fig:meanmodelbins} for $1.0^\circ$ pixels, which are the ones used for our analysis. We see that for the inner $9^\circ$, the model in each energy bin tends to underpredict the data, particularly in the highest energy bin. The overprediction of the model at radii larger than $9^\circ$ is seen to be mainly due to the lower-energy $\gamma$ rays.

The fit in \fig{fig:meanmodels} can be significantly improved by renormalizing the components. The model of inner $60^\circ$ profile is much improved if we increase the bremsstrahlung by 10\%, decrease the inverse Compton scattering by 60\%, and reduce the isotropic background to $6\times10^{-8}$ ph cm$^{-2}$s$^{-1}$sr$^{-1}$. However, such a renormalization procedure must also be accompanied by significant modifications of the spectrum of each component in order to account for different residuals in each energy bin shown in \fig{fig:meanmodelbins}. If such a renormalization is needed, it should appear as being needed in the GI profile, which we now investigate.

\subsection{GI-Profiles of Models}
\label{ssec:gimodel}
We now consider the consistency of the model with the GI component of the data. The emission in each annulus is dominated by emission in the Galactic disk, bright point sources, and nearby gas clouds obscuring the line-of-sight at higher Galactic latitudes. By considering the GI profile, we remove these bright features in an uncomplicated way.

\begin{figure*}
  \includegraphics[width=\columnwidth]{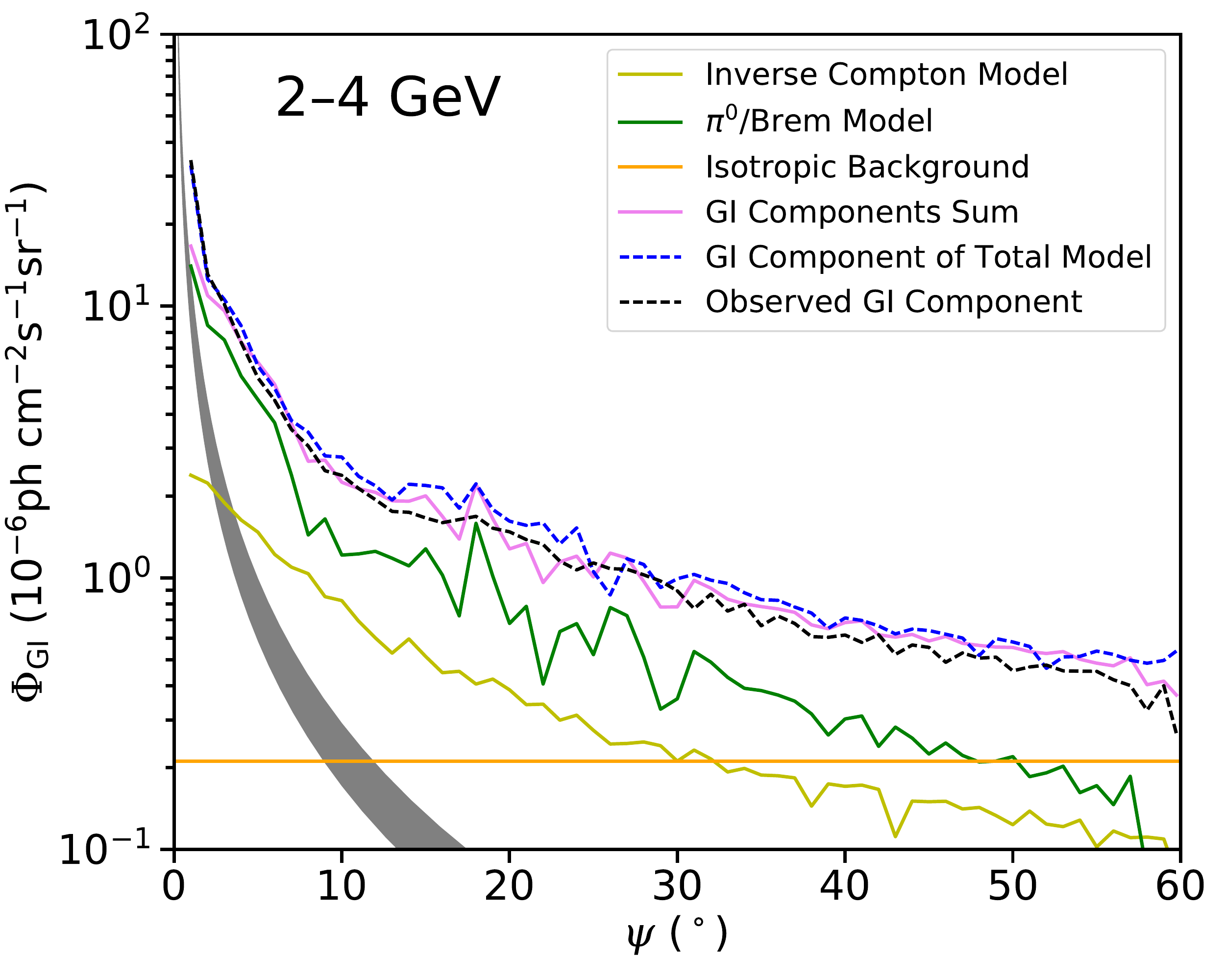}
  \includegraphics[width=\columnwidth]{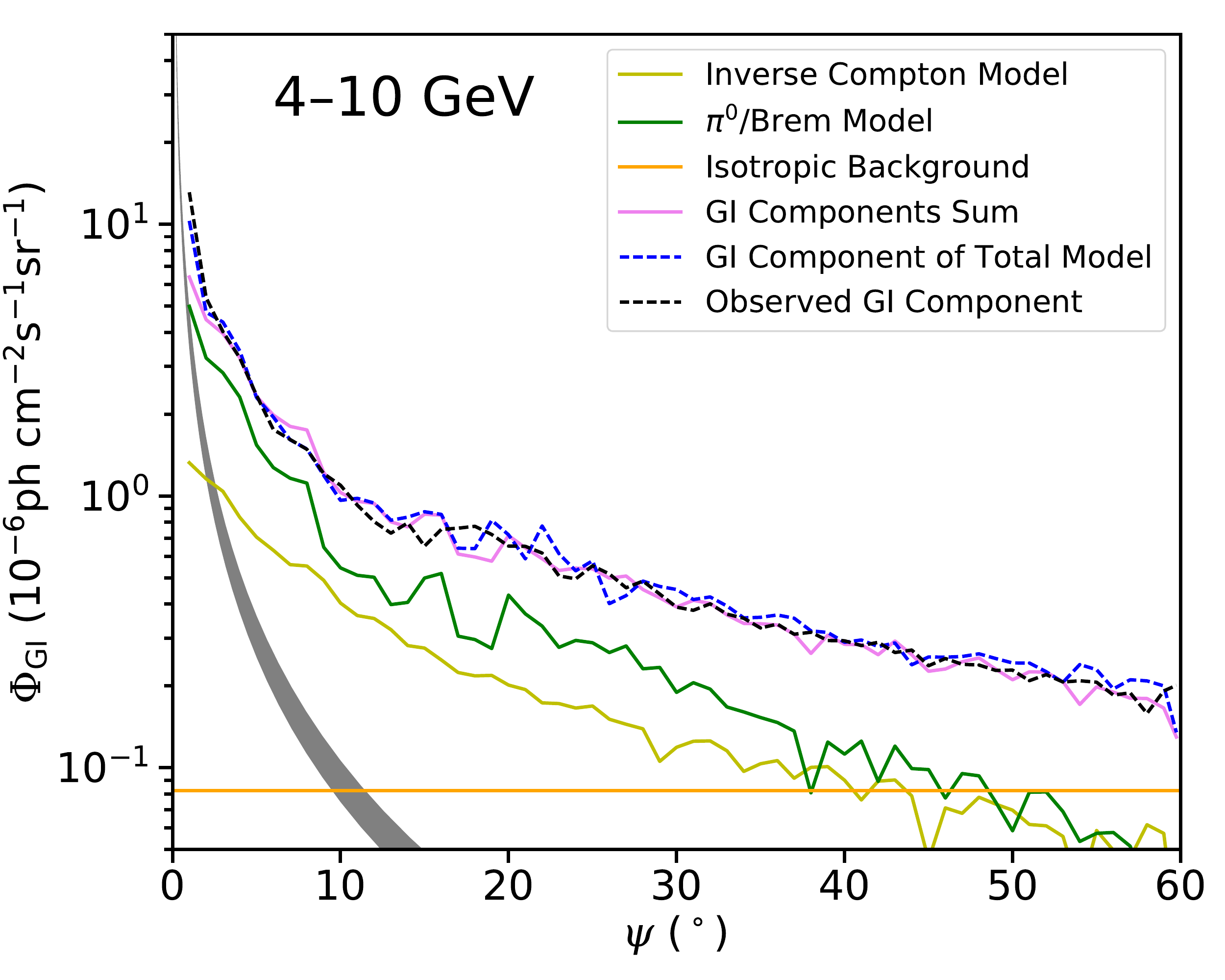}\\
  \includegraphics[width=\columnwidth]{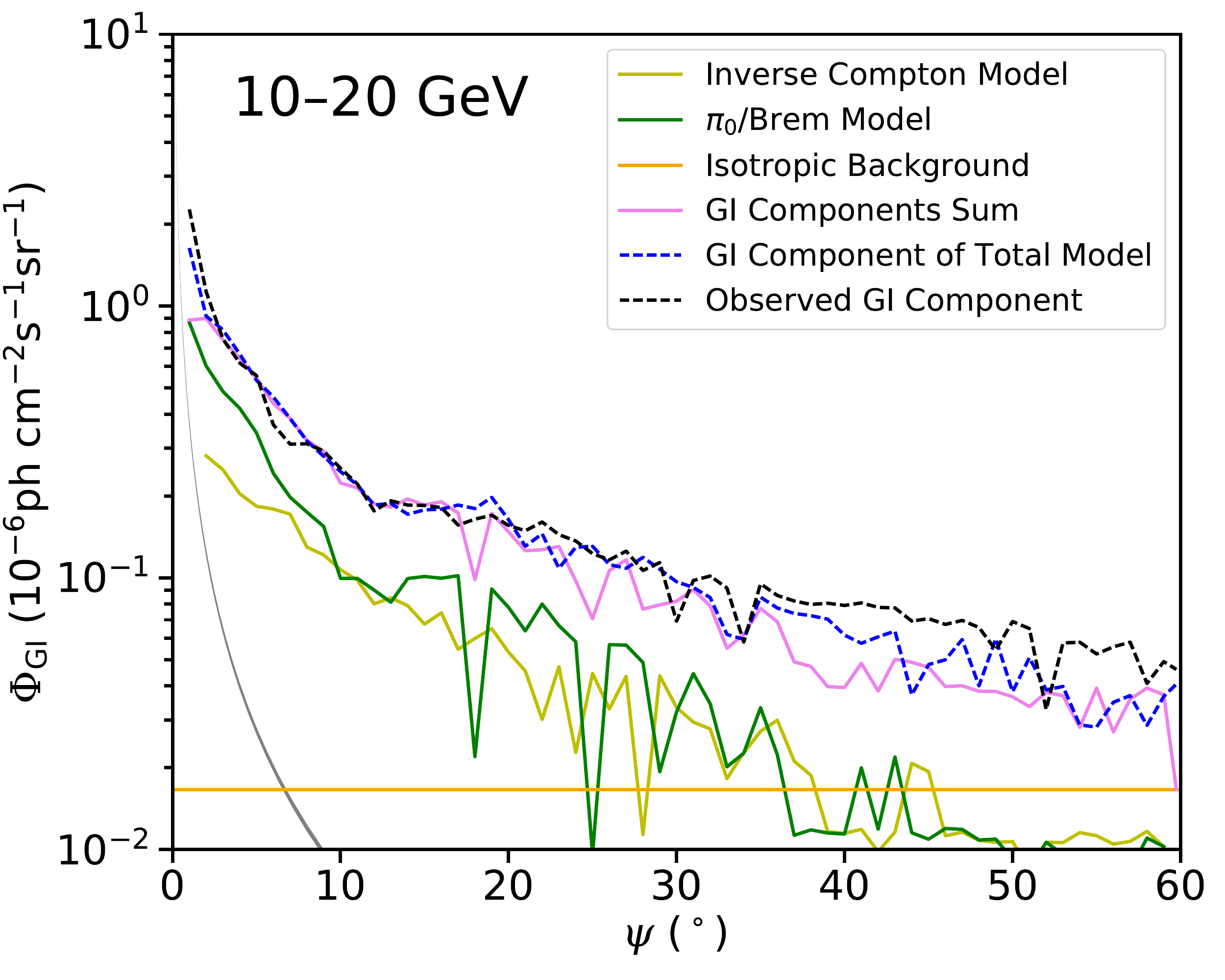}
  \includegraphics[width=\columnwidth]{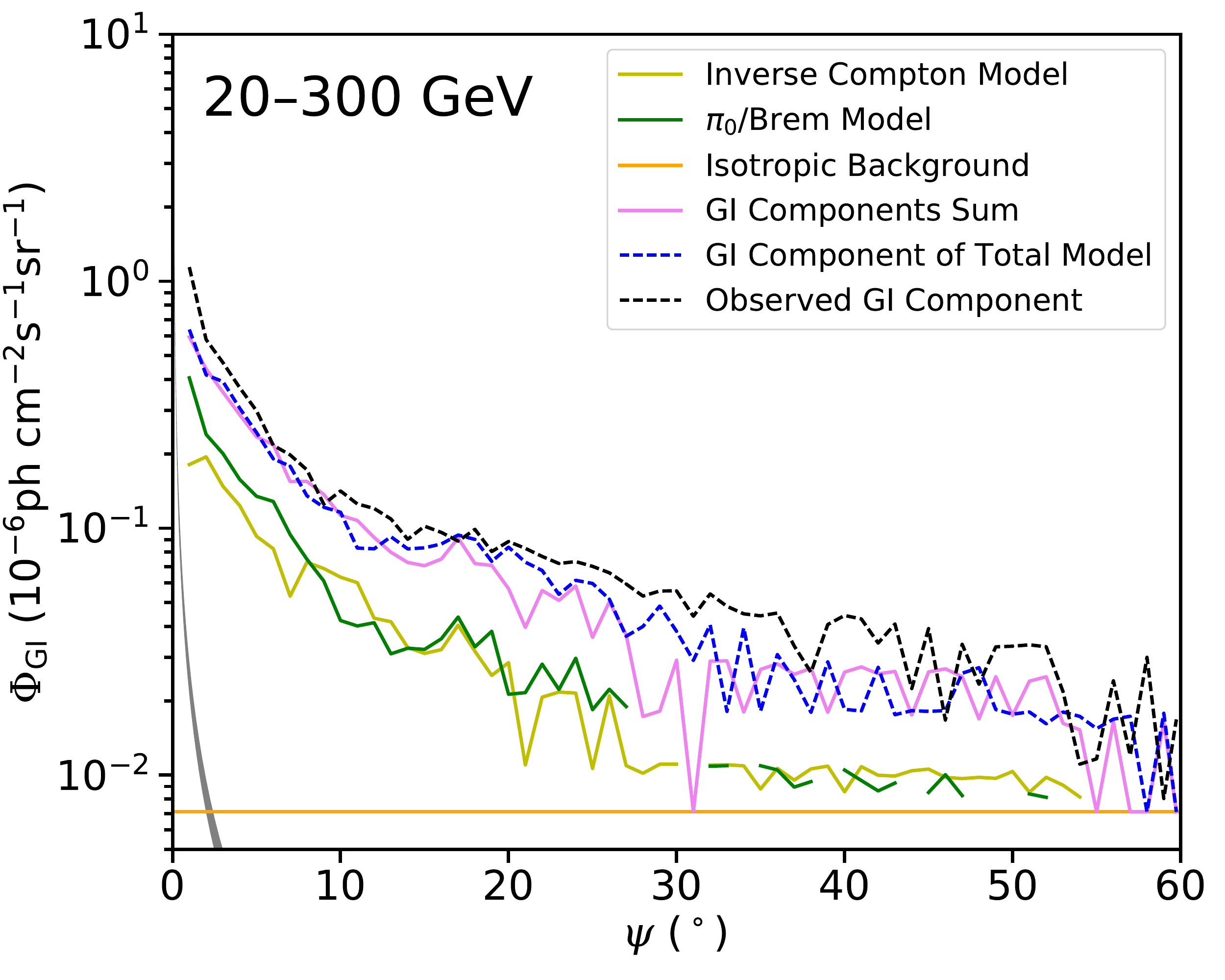}
  \caption{\label{fig:modelGI} The GI component (via the BDS test) of the observed $\gamma$ rays, the sky model, and each component of the model. At high energies and high radii, we see breaks in the lines where the BDS test failed to detect a GI flux. The differences between observations and model noted in \fig{fig:meanmodelbins} are not apparent in the GI flux, and must therefore be due to inaccuracies in the non-GI component. The spectrum of the Galactic center excess consistent with our halo and foreground model for annihilation to $b\overline{b}$ is shown by the grey bands.}
\end{figure*}

We begin by estimating the GI profile of each of our modeled sources to see how they compare to the GI profile of the data in each of our energy bins. These are shown in \fig{fig:modelGI}.

The only GI component of the 3FGL point sources is due to a few point sources clustered at the Galactic center. While this contributes to the GI component of the total model, it is not included in the sum of the GI fluxes for the model components.

The distribution of inverse Compton (IC) scattering is dominated by a thick disk that envelops the Galactic disk. The emission is also quite uniform in the inner regions of the disk. The thickness and uniformity of the feature makes most $0\fdg2$ annuli within the inner $5\fdg5$ consistent with being purely GI. The IC scattering is clearly an important part of the GI component of the inner Galactic region.

In comparison, the contribution of bremsstrahlung and $\pi^0$ production by cosmic rays in gas clouds forms a much thinner disk feature, but also more extended correlated structures outside the disk. This component is much more reduced by considering only its GI part. Nevertheless, it is a significantly brighter component than the IC scattering, and it remains the dominant component of the GI flux below 10 GeV.

Immediately we see in \fig{fig:modelGI} that the GI flux of the model (black, dashed curve) appears to provide remarkably good fidelity to the observed GI flux. The greatest discrepancy is found at $>30^\circ$ radii in the $>4$ GeV energy bins, where the model is consistently slightly below observations. At these large energies, there are fewer numbers of events particularly at higher latitudes, thus we can expect the BDS test to be unstable and underestimate the actual GI flux in this regime.

The agreement between the data and model shows that the suggested renormalization of components in \fig{fig:meanmodelbins} described in the previous subsection is not justified, since it would result in the modeled GI profile underestimating the data. We conclude that the discrepancies in \fig{fig:meanmodelbins} are due to mismodeling of structure, in the non-GI component. This example illustrates the usefulness of the GI component to provide useful information that can be used to improve full-sky models.

One final result of interest is that, mathematically speaking, it is not strictly true that the sum of GI components is the GI component of the total. Low contrast correlated structure in $\pi^0$ emission could become obscured when combined with IC radiation. In addition, higher resolution structure could be visible in the total model that is not in the components because the fewer counts per pixel in the components may not support the stability of the BDS test at higher resolutions. In \fig{fig:modelGI}, we compare the sum of estimated GI components of each model component to the estimated GI component of the total model in each energy bin. Their remarkable consistency suggests that it is valid to interpret the GI flux of the model as the sum of the GI flux of each model component.

\section{Dark Matter Constraints}
\label{sec:dm}

By measuring and understanding the GI component of the $\gamma$-ray sky, we robustly increase the signal-to-noise of potential exotic emission that has a strong GI component. An obvious example is dark matter annihilation or decay. The GeV Galactic center excess is expected to be predominantly GI; hence, it can be detected and characterized in a GI profile analysis, which could further restrict its possible interpretations.

The radiation from a dark matter signal would have a distribution related to the density profile of the Milky Way dark matter halo and its substructure. While large subhalos and correlated substructures such as tidal streams would be non-GI components, emission from the smooth halo profile and evenly distributed unresolvable subhalos are expected to be the dominant features of a dark matter signature. Since the gamma-ray anisotropy at high galactic latitudes does not currently show an indication of degree-scale contributions of a dark matter signal to correlated structure \cite{Fornasa:2016ohl}, it severely constrains a non-GI contribution of dark matter to the data.

This work provides the first attempt to constrain dark matter by limiting its predicted dominant feature, its GI component, against a conservative model of the diffuse $\gamma$-ray sky. 

The dark matter halo profile of the Milky Way halo is taken for this analysis to be a generalized Navarro-Frenk-White (NFW) profile
\begin{equation}
  \rho(r)=\rho_\odot\left(\frac{R_\odot}{r_s}\right)^{\!\!\gamma}\left(1+\frac{R_\odot}{r_s}\right)^{\!\!3-\gamma}\left(\frac{r}{r_s}\right)^{\!\!-\gamma}\left(1+\frac{r}{r_s}\right)^{\!\!-(3-\gamma)}
\end{equation}
as used in \citet{Calore:2014nla} with inner slope $\gamma=1.26$, scale radius $r_s=20$ kpc, where the Sun's distance from the Galactic center is $R_\odot=8.5$ kpc, and the dark matter density at the Solar System is $\rho_\odot=0.4$ GeV cm$^{-3}$. Such a profile was chosen to be consistent with the profile of $\gamma$ rays near the Galactic center associated with the GeV excess \citep{TheFermi-LAT:2015kwa}.

\begin{figure}
  \includegraphics[width=\columnwidth]{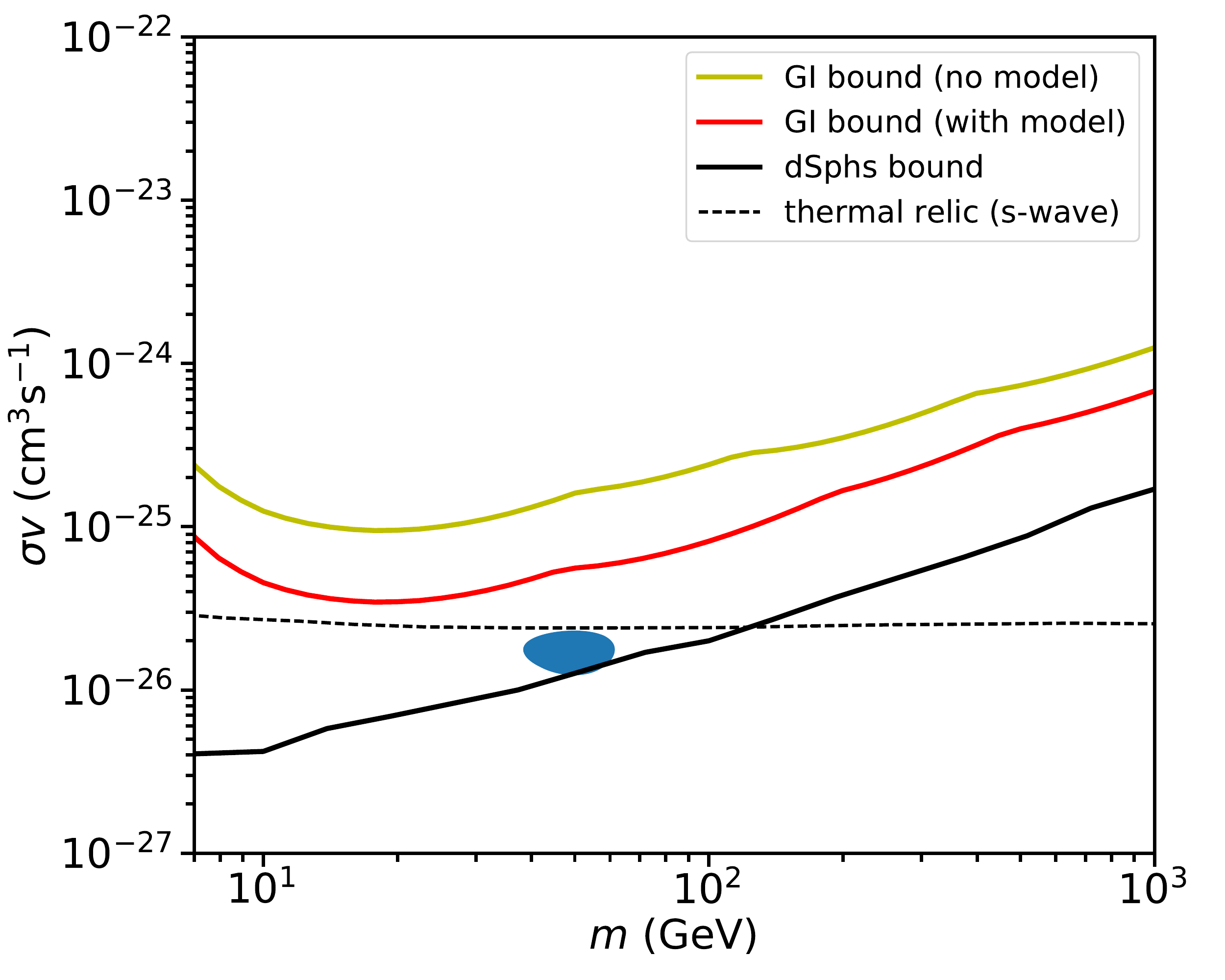}
  \caption{\label{fig:dmlimits} GI $\gamma$-ray flux limits on dark matter annihilation (to $b\overline{b}$ quarks) with and without modeling the GI flux, compared with the most recent dwarf satellite limits from \emph{Fermi}-LAT and \emph{MAGIC}. Constraints from the GI flux are very close to the thermal s-wave annihilating relic cross-section \citep{Steigman:2012nb}. The Milky Way halo model is described in the text.}
\end{figure}

In \fig{fig:dmlimits}, the blue ellipse indicates the approximate range of dark matter particle mass $m$ and velocity-weighted annihilation cross-section $\sigma v$ of dark matter annihilating to $b\overline{b}$ particles that is consistent with the GeV excess for the model A background model in \citet{Calore:2014nla}, which we are using. We see that such a model is in tension with the most current bound from the dwarf spheroidal galaxies, indicated by the black line \citep{Ahnen:2016qkx}, as has been pointed out more generally \citep{Abazajian:2015raa}. For comparison, the GI profile of this model of the GeV excess signal is indicated in \fig{fig:fluxplot} and \fig{fig:modelGI} where we see that it makes up a small fraction of the GI profile in the inner Galaxy, but is certainly observable within $10^\circ$ of the Galactic center, once the GI flux can be determined to the needed precision.

The allowed flux from possible exotic sources is taken as the difference of a conservatively underestimated modeled flux from a conservatively overestimated observed flux.

As explained in \ssec{ssec:gimodel}, the conservative observed flux in each of our energy bins is taken to be the GI flux determined by the BDS test, as shown by the gray lines for each energy bin in \fig{fig:modelGI}. The radial range of $8-20^\circ$ is the range most important to our constraints.

The most conservative model is to use no diffuse background model at all. The dark matter constraint for this case is indicated by the top (yellow) curve in \fig{fig:dmlimits}. Even this most conservative of limits comes quite close to the thermal relic cross-section for s-wave annihilating dark matter, indicated by the thin dashed line in the figure. However, there are clear contributions to the GI flux by inverse Compton scattering, bremsstrahlung, and the isotropic background. A very conservative estimate of their contribution is found by taking the GI component of the total $\gamma$-ray sky model (the dashed black curves in \fig{fig:modelGI}), and reducing it by $30\%$, the estimated uncertainty of the BDS test's estimate of the GI flux. Subtracting this foreground contribution from the observed GI flux and not allowing dark matter annihilation to overproduce the remaining flux reduces the constraint to the middle (red) curve in \fig{fig:dmlimits}.

\section{Summary and Outlook}
\label{sec:discussion}
We have explored methods for extracting the galacto-isotropic (GI) component of astrophysical $\gamma$ rays, and carried out a preliminary analysis on 85 months of Fermi-LAT data. The results of this analysis provide a number of insights useful for interpreting the $\gamma$-ray sky.

We considered two types of non-GI contaminants: non-Poissonities (NPs) and spatial correlations (SCs) within the pixels of a fixed annulus about the Galactic center, in a GI-tiling. Both types are observed to be nearly identical to each other, which is consistent with the standard picture that the non-GI contributions to the $\gamma$ skymap are predominantly due to point sources and cosmic rays interacting with giant gas clouds. After removing the brightest pixels where these effects appear, the remaining dim pixels are consistent with being structureless and Poisson distributed. While it is ideal that a determination of the GI profile removes both NPs and SCs simultaneously rather than just one, the resulting gains will be marginal.

A BDS analysis is used to remove SCs, without consideration as to whether the uncorrelated pixels are Poisson-distributed. The BDS method is mature and efficient, but is unstable if the mean number of events per pixel is too low, or if the number of pixels in the annulus is too small. In addition, a precise estimate of the uncertainty in the resulting GI profile requires Monte-Carlo simulations of mock $\gamma$-ray data from model skymaps.

We introduced the ordered-Poisson-ensemble, fit to the dim pixels in an annulus, to detect the presence of NPs. It does not consider if the dim, Poisson-distributed pixels are spatially correlated. This method is amenable to analytic methods. Hence, further development of the ordered-Poisson method has the potential to analytically determine the statistical uncertainty of the GI profile. This method is also more stable than the BDS analysis, being applicable for any number of pixels containing any distribution of events.

The GI-profile estimate is remarkably consistent across these independent methods (and within their variations), indicating the robustness of estimating the GI profile. This variance is used as a rough systematic estimate of the GI profile uncertainty in this work.

We provided an interpretation of the GI flux in terms of the GI components of a full-sky model containing bremsstrahlung and $\pi^0$ emission, inverse Compton scattering of starlight, the 3FGL point source catalog, and the isotropic background. We used a pre-defined model that is observed to systematically overestimate the observed flux at galactocentric radii larger than $\sim10^\circ$. It is surprising how remarkably well the GI component of the model matches the observed GI flux. This means that the overestimated flux occurs in the structured part of the sky, such as the Galactic plane. This provides important new constraints for refining models of diffuse $\gamma$-ray sources.

Another useful result is that the GI component of the sky model is found to consistently decompose into the GI components of each of the distinct sources. This allows us to cleanly assign the amount of the GI flux that is coming from each source.

If dark matter annihilation were to contribute to the $\gamma$-ray sky, the GI flux would contain all dark matter emission except that coming from resolvable substructure of the Milky Way, or resolvable extragalactic structure in the local Universe. For the model of the Milky Way dark matter density profile chosen to fit the GeV excess in the Galactic center, the measured GI profile limits the dark matter annihilation cross section to $\sigma v<10^{-25}$ cm$^3$ s$^{-1}$ for a particle mass of 20 GeV annihilating to $b\overline{b}$. After accounting for the cosmic ray and point-source emission, this limit is improved to $\sigma v<4\times10^{-26}$ cm$^3$ s$^{-1}$ at the current level of systematic uncertainty.

We have not yet measured the GI profile to the level of precision needed to observe the GeV excess. Yet, the excess is observed to high significance using likelihood fits of the Fermi-LAT data with the same sky model we applied. If the excess is galacto-isotropic, there is no reason why the GI methods should not also be able to detect it with the same level of significance. Thus, we expect the GI profile to be measured with a precision of $\sim10\%$ with the present Fermi-LAT data. This will be accomplished with more detailed statistical studies of the GI flux estimator; for example, by measuring its statistical distribution with Monte-Carlo simulations.

If, at this refined precision, the GeV excess is still not detected, then it must have a significant non-GI component. Indeed, non-Poissonities have been identified in the Galactic center excess \citep{Bartels:2015aea,Lee:2015fea,Fermi-LAT:2017yoi}, but the flux of the underlying GI component has not yet been quantified.

The ordered-Poisson-ensemble analysis has the potential to be able to determine the GI flux with an exact likelihood method. This would bring the precision of the GI-flux measurement to the level of Poisson noise and instrument systematics, as well as contributions from residual spatial correlations in the data after non-Poissionities are removed. If remaining spatial correlations become restrictive, then new GI flux estimators that remove these structures could further improve precision.

\section*{Acknowledgments}
MK \& SSC are supported by National Science Foundation Grant PHY-1620638. AK is supported by NSF GRFP Grant DGE-1321846. SSC acknowledges support from a McCue Fellowship.

\bibliographystyle{mnras}
\bibliography{bds_analysis}

\appendix

\section{Higher precision GI Maps}
\label{app:highprecgi}
\begin{figure*}
	\includegraphics[width=\linewidth]{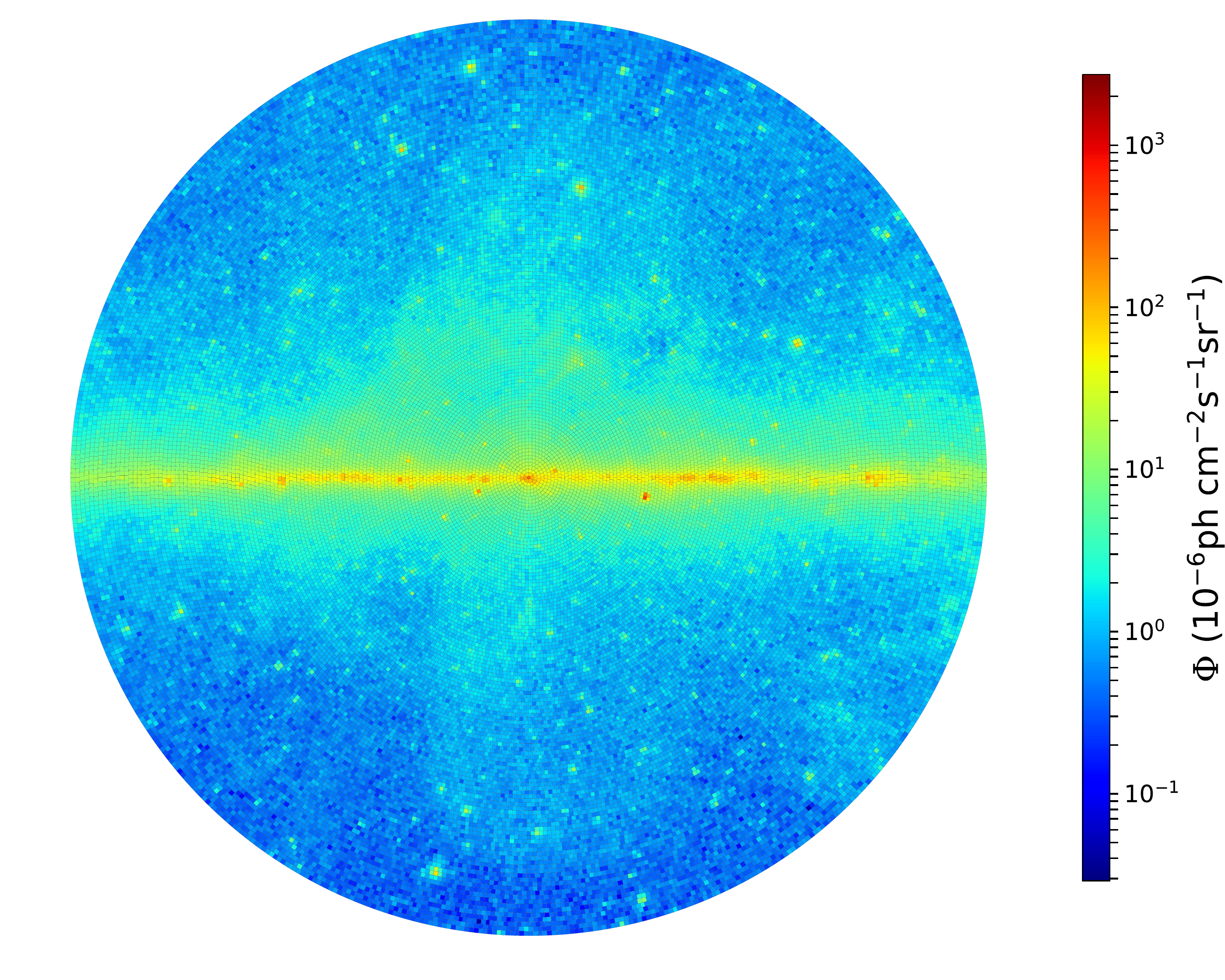}
	\caption{\label{fig:GItiling2}Same as \fig{fig:GItiling1} except at $0\fdg5$ resolution.}
\end{figure*}
\begin{figure*}
	\includegraphics[width=\linewidth]{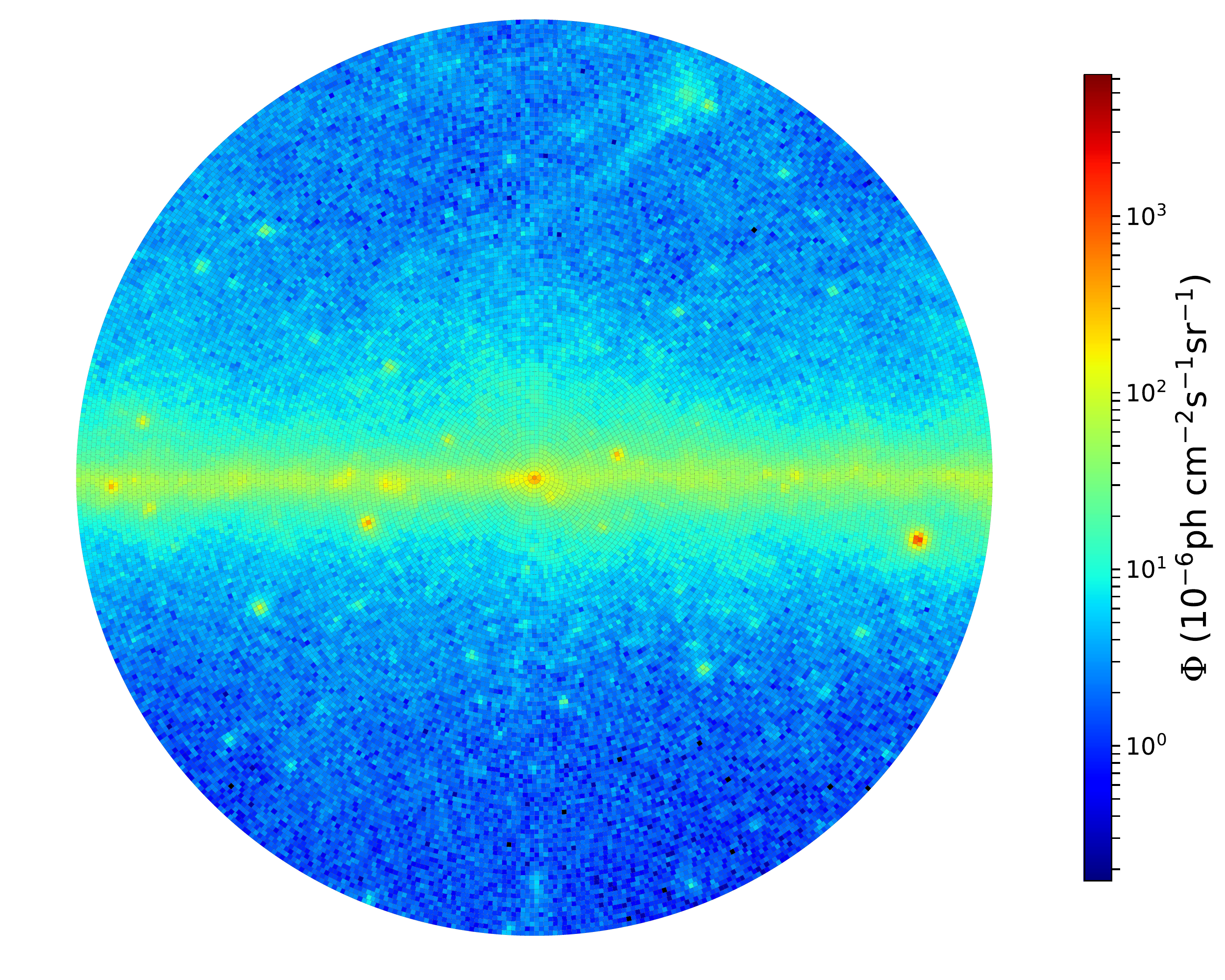}
	\caption{\label{fig:GItiling3}Same as \fig{fig:GItiling1} except at $0\fdg2$ resolution out to $20^\circ$ from the Galactic center.}
\end{figure*}
\begin{figure*}
  \includegraphics[width=0.83\linewidth]{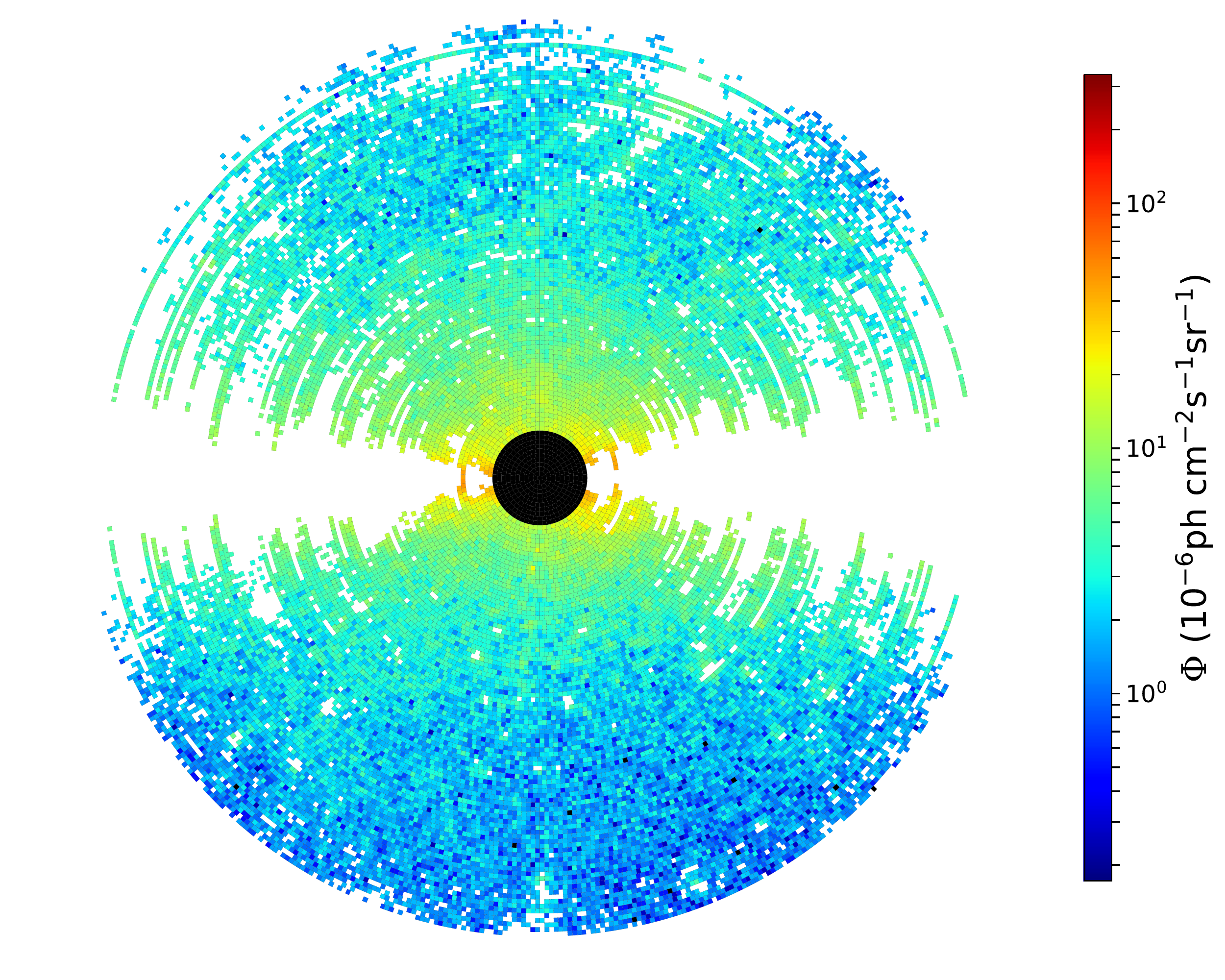}
  \caption{\label{fig:BDStest3}The result of using the BDS test with interleaving pixels to remove structure on the map in Fig.~\ref{fig:GItiling3}.}
\end{figure*}

Figs.~\ref{fig:GItiling2} and \ref{fig:GItiling3} show portions of the GI maps at sub-degree resolutions. We show in Fig.~\ref{fig:BDStest3} the result of applying our BDS test to remove spatial correlations in the highest-resolution data. Visually, we can see that the removal of point sources and structured clouds is efficiently removed, and the radial GI profile has clearly emerged.

\section{Statistical width of the BDS correlator}
\label{app:BDSsigma}
The variance of the BDS correlation integral $C(\varepsilon,m,N)$ is estimated by
\begin{align*}
  &\sigma^2\!(\varepsilon,m,N)=4\big[k_N\!(\varepsilon)\!-\!c_N^2\!(\varepsilon)\big]^2\sum_{n=1}^{m-1}n^2k_N^{m-n-1}\!(\varepsilon)\big[c_N^2\!(\varepsilon)\big]^{n-1}\\
  &\quad=4\Big[k_N^m\!(\varepsilon)+2\sum_{n=1}^{m-1}k_N^{m-n}\!(\varepsilon)c_N^{2n}\!(\varepsilon)+(m-1)^2c_N^{2m}\!(\varepsilon)\\
  &\qquad-m^2k_N\!(\varepsilon)c_N^{2m-2}\!(\varepsilon)\Big]
\end{align*}
for the ensemble mean scalar 2-point correlator
\begin{equation*}
  c_N(\varepsilon)\equiv\mean{C(\varepsilon,1,N)}\approx C(\varepsilon,1,N),
\end{equation*}
well-estimated by the sample's scalar correlator, and the 3-point correlator
\begin{equation*}
  k_N\!(\varepsilon)=\frac{6}{N(N-1)(N-2)}\sum_{i_1=0}^{N-1}\,\sum_{i_2=i_1+1}^{N-1}\,\sum_{i_3=i_2+1}^{N-1}\!\!\!b_\varepsilon\!(\Phi^m_{i_1},\Phi^m_{i_2},\Phi^m_{i_3})
\end{equation*}
with
\begin{equation*}
  b_{\varepsilon\,}\!(1,2,3)=\frac{1}{3}\big[I_{\hspace{-1pt}\varepsilon\,}\!(1,2)\,I_{\hspace{-1pt}\varepsilon\,}\!(2,3)+I_{\hspace{-1pt}\varepsilon\,}\!(1,3)\,I_{\hspace{-1pt}\varepsilon\,}\!(3,2)+I_{\hspace{-1pt}\varepsilon\,}\!(2,1)\,I_{\hspace{-1pt}\varepsilon\,}\!(1,3)\big].
\end{equation*}

\section{Ordered Poisson Statistics}
\subsection{Unconditional Mean Ordered Poisson Ensemble}
\label{app:aveordered}

Consider an ensemble of $N_{\text{pix}}$ pixels, each with the same constant intensity $I$, and observed with the same exposure $\varepsilon$. The expectation count of each pixel is $n_{\text{exp}}=I\varepsilon$, but the observed count in each pixel $n_i$ is a random variable with probability distribution given by the Poisson distribution
\begin{equation}
  \label{eqn:poisson}
  P_P(n|n_{\text{exp}})=e^{-n_{\text{exp}}}\frac{n_{\text{exp}}^n}{n!}.
\end{equation}
We are interested in the distribution of the flux-ordered ensemble $n_i$ vs. $i$ when $n_1\leq n_2\leq\cdots\leq n_{N_\text{pix}}$.

To begin, we determine the probability that the $i^{\text{th}}$ dimmest pixel has $n_i=n$ counts. In this subsection, we focus on the unconditional probability that is irrespective of the counts in the other pixels. We delay until \app{app:condordered} considering the conditional probability of $n_i$ given the value of the dimmer pixel counts.

Our strategy is to sum over the probabilities of each configuration where any pixel has $n$ counts, $i-1$ pixels have $\leq n$ counts, and $N_{\text{pix}}-i$ pixels have $\geq n$ counts. Define $N_n$ as the number of pixels with precisely $n$ counts, $N_-$ as the number of pixels with fewer than $n$ counts, and $N_+=N_{\text{pix}}-N_n-N_-$ as the number of pixels with more than $n$ counts.

It is useful to define the pixel probability to have a count less than $n$ as
\begin{equation}
  P_-(n|n_{\text{exp}})\equiv\sum_{m=0}^{n-1}P_P(m|n_{\text{exp}}),
\end{equation}
and similarly for a count greater than $n$,
\begin{align}
  P_+(n|n_{\text{exp}})&\equiv\sum_{m=n+1}^{\infty}P_P(m|n_{\text{exp}})\\
  &=1-P_P(n|n_{\text{exp}})-P_-(n|n_{\text{exp}})\\
  &=1-P_-(n+1|n_{\text{exp}}).\nonumber
\end{align}

The probability that $n_i=n$ can then be written as
\begin{align}
  P(n_i=n)=&\sum_{N_n=1}^{N_{\text{pix}}}\sum_{N_-=\max(0,i-N_n)}^{\min(i-1,N_{\text{pix}}-N_n)}\binom{N_{\text{pix}}}{N_n}\binom{N_{\text{pix}}-N_n}{N_-}\nonumber\\
  &[P_-(n)]^{N_-}[P_P(n)]^{N_n}[P_+(n)]^{N_{\text{pix}}-N_n-N_-},
\end{align}
where $\binom{\cdot}{\cdot}$ is the binomial coefficient. Understanding this result involves the following considerations. The range of values of $N_n$ can be anywhere from only $1$ pixel to all $N_{\text{pix}}$ pixels. For fixed $N_n$, all possible sets of $N_n$ pixels represents a unique configuration. The number of such sets is $\binom{N_{\text{pix}}}{N_n}$. Then, the lowest value that $N_-$ can have is $i-N_n$. This is the configuration with all $N_n$ pixels with $n$ counts occur within the first $i$ dimmest pixels. If $N_n>i$, then the smallest value of $N_-$ is $0$. The largest value that $N_-$ can have is $N_{\text{pix}}-N_n$ for large $N_n$. This is the configuration where all of the $N_{\text{pix}}-i-1$ brightest pixels have $n$ counts. If $N_n<N_{\text{pix}}-i$, then the maximum value of $N_-$ is $i-1$, the possible number of pixels dimmer than the $i^{\text{th}}$ dimmest. After choosing $N_n$ pixels that have $n$ events, we get a new configuration for every set of $N_-$ pixels from the remaining $N_{\text{pix}}-N_n$. The number of such sets is $\binom{N_{\text{pix}}-N_n}{N_-}$.

Using the binomial theorem, it is straightforward to arrive at the alternative expression
\begin{align}
&P(n_i=n)=1-\Big(1-P_P(n)\Big)^{\!N_{\text{pix}}}\nonumber\\
	&\ \  -\left[\sum_{N_n=1}^{N_{\text{pix}}-i}\ \sum_{N_-=i}^{N_{\text{pix}}-N_n}+\sum_{N_n=1}^{i-1}\sum_{N_-=0}^{i-N_n-1}\right]\binom{N_{\text{pix}}}{N_n}\binom{N_{\text{pix}}-N_n}{N_-}\nonumber\\
  &\qquad[P_-(n)]^{N_-}[P_P(n)]^{N_n}[P_+(n)]^{N_{\text{pix}}-N_n-N_-}.
\end{align}
This expression takes the alternative approach of subtracting from $1$ the probabilities of all configurations that do not have $N_n$ pixels with $n$ counts. The second term removes those configurations where no pixels have $n$ counts. The first set of sums removes configurations with too many pixels with fewer counts than $n$, i.e., $N_-$ is too large than is possible with the given value of $N_n$. The second set of sums removes those configurations where $N_-$ is too small.

Given the unconditional probabilities for the $i^{\text{th}}$ dimmest pixel count, the probabilities for the $(i+1)^{\text{th}}$ dimmest pixel can be determined recursively. Start with the dimmest pixel $i=1$.
\begin{equation}
  P(n_1=n)=[1-P_-(n)]^{N_{\text{pix}}}-[P_+(n)]^{N_{\text{pix}}}
\end{equation}
The unconditional count probability for brighter pixels can be determined with
\begin{align}
  P(n_{i+1}=n)&=P(n_i=n)\\
  &\quad+\binom{N_{\text{pix}}}{i}\Bigg\{[P_-(n)]^i[1-P_-(n)]^{N_{\text{pix}}-i}\nonumber\\
  &\hspace{60pt}-[1-P_+(n)]^i[P_+(n)]^{N_{\text{pix}}-i}\Big\}\nonumber
\end{align}

The unconditional mean counts in this paper were determined for a given value of $n_{\text{exp}}$ with the following algorithm.
\begin{enumerate}
\item Compute a look-up table of $P_+(n)$ for $n=0,1,2,\ldots,n_{\text{max}}$ with $n_{\text{max}}$ chosen sufficiently large that $P_+(n_{\text{max}})<\delta$ for $\delta$ negligible for the desired precision. For our examples, we used $\delta=10^{-8}$.
\item The average count of the dimmest pixel is
\begin{equation}
  \mean{n_1}=\sum_{n=1}^{n_{\text{max}}}n\Big\{[P_+(n-1)]^{N_{\text{pix}}}-[P_+(n)]^{N_{\text{pix}}}\Big\}.
\end{equation}
\item The mean counts of subsequently brighter pixels are determined recursively.
\begin{align}
  \mean{n_{i+1}}&=\mean{n_i}\\
  &+\binom{N_{\text{pix}}}{i}\sum_{n=1}^{n_{\text{max}}}n\Big\{[1-P_+(n-1)]^i[P_+(n-1)]^{N_{\text{pix}}-i}\nonumber\\
  &\hspace{75pt}-[1-P_+(n)]^i[P_+(n)]^{N_{\text{pix}}-i}\Big\}\nonumber
\end{align}
\item We verify the precision of the brightest pixel determined recursively by comparing the result with the precise expression
\begin{equation}
  \mean{n_{N_{\text{pix}}}}=\sum_nn\Big\{[1-P_+(n)]^{N_{\text{pix}}}-[1-P_+(n-1)]^{N_{\text{pix}}}\Big\}.
\end{equation}
\end{enumerate}

\subsection{Conditional Mean Ordered Poisson Ensemble}
\label{app:condordered}

There are a number of different conditional moments that can be considered for ordered ensembles. For demonstration, we consider the specific case of the probability of the flux in the $i^{\text{th}}$ dimmest pixel given the fluxes of every dimmer pixel, $P(n_i|n_1,\ldots,n_{i-1})$, with $n_1\leq\cdots\leq n_{i-1}$.

We use the same strategy as before, by adding up the probability of all remaining allowed ensembles. With the fluxes of $i-1$ pixels known, the counts of the remaining $N_{\text{pix}}-i+1$ pixels must be at least $n_{i-1}$. In this situation, the Poisson probabilities of interest are
\begin{equation}
  P(n|n\geq n_{i-1})=\frac{P_P(n)}{P_+(n_{i-1}-1)}.
\end{equation}

Let $N_n$ be the number of the remaining $N_{\text{pix}}-i+1$ pixels that have the same count as $n_i$. The range of $N_n$ is $1\leq N_n\leq N_{\text{pix}}-i+1$. For a fixed value of $N_n$, there are $\binom{N_{\text{pix}}-i+1}{N_n}$ possible ways to choose the $N_n$ pixels, after which the remaining $N_{\text{pix}}-i-N_n+1$ pixels can have any values greater than $n_i$.
\begin{align}
  \label{eq:condprob}
  &P(n_i|n_1\leq\cdots\leq n_{i-1})=\sum_{N_n=1}^{N_{\text{pix}}-i+1}\binom{N_{\text{pix}}-i+1}{N_n}\\
  &\qquad\qquad\left[\frac{P_P(n_i)}{P_+(n_{i-1}-1)}\right]^{N_n}\left[\frac{P_+(n_i)}{P_+(n_{i-1}-1)}\right]^{N_{\text{pix}}-i+1-N_n}\nonumber\\
  &=\left[\frac{P_+(n_i-1)}{P_+(n_{i-1}-1)}\right]^{N_{\text{pix}}-i+1}-\left[\frac{P_+(n_i)}{P_+(n_{i-1}-1)}\right]^{N_{\text{pix}}-i+1}\nonumber
\end{align}
This result can be applied to determining likelihood functions for ordered Poisson ensembles.

\subsection{Upper Threshold Likelihood Function for Ordered Poisson Ensembles}
\label{app:likelihood}

One possible method for determining the GI flux in an annulus is to determine the flux from $n_{\text{exp}}$ that maximizes the ordered Poisson likelihood function for some threshold number of the dimmest pixels $N_{\text{th}}$. We want to choose $N_{\text{th}}$ as large as is consistent with that many dimmest pixels being uncontaminated by structured foregrounds. For example, one could determine $n_{\text{exp}}$ for increasing values of $N_{\text{th}}$ until the goodness-of-fit of the dimmest $N_{\text{th}}$ pixels of the mean ordered-Poisson ensemble to the data becomes worse than some allowed tolerance.

For a given value of $N_{\text{pix}}$ and $N_{\text{th}}\leq N_{\text{pix}}$, the likelihood function can be determined from conditional probabilities.
\begin{align}
  &L(n_{\text{exp}}|N_{\text{th}},N_{\text{pix}},n_1,n_2,\ldots,n_{N_{\text{th}}})\\
  &\qquad=P(n_1\,\&\,n_2\,\&\cdots\&\,n_{N_{\text{th}}}|n_{\text{exp}},N_{\text{th}},N_{\text{pix}})\nonumber
\end{align}
In terms of the conditional probabilities in \eqn{eq:condprob},
\begin{align}
  &L(n_{\text{exp}}|n_1,n_2,\ldots,n_{N_{\text{th}}})\\
  &\qquad=P(n_1)P(n_2|n_1)\cdots P(n_{N_{\text{th}}}|n_1,n_2,\ldots,n_{N_{\text{th}}-1})\nonumber\\
  &\qquad=\frac{\prod_{i=1}^{N_{\text{th}}}\Big[P_+^{N_{\text{pix}}-i+1}(n_i-1)-P_+^{N_{\text{pix}}-i+1}(n_i)\Big]}{\prod_{j=1}^{N_{\text{th}}-1}P_+^{N_{\text{pix}}-j}(n_j-1)}\nonumber\\
  &\qquad=\left\{\prod_{i=1}^{N_{\text{th}}}P_+(n_i-1)\left[1-\left(\frac{P_+(n_i)}{P_+(n_i-1)}\right)^{\!\!N_{\text{pix}}-i+1}\right]\right\}\nonumber\\
  &\qquad\quad\  P_+^{N_{\text{pix}}-N_{\text{th}}}(n_{N_{\text{th}}}-1).\nonumber
\end{align}

The log-likelihood function is efficiently computed using loggamma functions. First, note that the lower Poisson probabilities are upper incomplete Gamma functions
\begin{equation}
  P_+(n|n_{\text{exp}})=\frac{\gamma(n+1,n_{\text{exp}})}{\Gamma(n+1)}=\frac{1}{n!}\int_0^{n_{\text{exp}}}dt\,t^{n-1}e^{-t}
\end{equation}
for non-negative $n$ (it is simply $1$ for negative $n$). Let $i_0$ be the lowest ordered-pixel index with $n_{i_0}>0$. We assume that $N_{\text{th}}>i_0$. The log-likelihood function is then computed with
\begin{align}
  &\ln L(n_{\text{exp}}|N_{\text{pix}},N_{\text{th}},n_1,\ldots,n_{N_{\text{th}}})\\
  &\qquad=(N_{\text{pix}}-N_{\text{th}})\ln\gamma(n_{N_{\text{th}}},n_{\text{exp}})\nonumber\\
    &\qquad\qquad+\sum_{i=1}^{i_0-1}\ln\Big[1-(1-e^{-n_{\text{exp}}})^{N_{\text{pix}-i+1}}\Big]\nonumber\\
  &\qquad\qquad+\sum_{i=i_0}^{N_{\text{th}}}\Bigg\{\ln\gamma(n_i,n_{\text{exp}})\nonumber\\
  &\qquad\qquad\qquad\quad+\ln\left[1-\left(\frac{\gamma(n_i+1,n_{\text{exp}})}{n_i\gamma(n_i,n_{\text{exp}})}\right)^{\!\!N_{\text{pix}}-i+1}\right]\Bigg\}\nonumber
\end{align}
up to an additive constant that is independent of $n_{\text{exp}}$.

\section{Model of the \emph{Fermi}-LAT Effective Point Spread Function}
\label{app:psf}
The instrument point spread function is estimated by the Fisher distribution,
\begin{equation}
  F(\theta)=\frac{1}{4\upi\sigma_b^2}\text{csch}(\sigma_b^{-2})\exp(\sigma_b^{-2}\cos\theta),
\end{equation}
which reduces for small $\sigma_b$ to the two-dimensional Gaussian in the flat sky approximation
\begin{equation}
  \label{eq:gaussianpsf}
  F(\theta)=\frac{1}{2\upi\sigma_b^2}\exp\left[-\frac{\theta^2}{2\sigma_b^2}\right].
\end{equation}
These are parametrised with an energy-dependent width $\sigma_b$, and $\theta$ is the angular distance from the source in radians. The $68\%$ containment angle for the SOURCE class of events is very well fit by \citep{Ackermann:2013uma}\footnote{The fitting function in \citet{Ackermann:2013uma} continues to be a good fit to the new Pass 8 instrument response functions, displayed at \url{http://www.slac.stanford.edu/exp/glast/groups/canda/lat_Performance.htm}, but with modified fit parameters.}
\begin{equation}
  \theta_{68}(E)=\frac{\upi}{180}\left[\left(0.81E^{-0.81}\right)^2 + 0.1^2\right]^{1/2}.
\end{equation}
The $68\%$ containment angle associated with \eqn{eq:gaussianpsf} is $\theta_{68}=\sqrt{\ln0.32^{-2}}\simeq1.51\sigma_b$. This is used to determine the effective width of each energy bin, and to populate each pixel of the GI tiling with an estimated flux from the 3FGL point sources in each energy bin\footnote{While the tails of the Fermi-LAT PSF are broader than Gaussian, this effect is not expected to be significant at the precision of the present study.}.

The cumulative point spread over an energy range $E_1<E<E_2$ when observing a flux spectrum $d\Phi/dE$ is
\begin{equation}
  F(\theta,E_1,E_2)=\frac{\int_{E_1}^{E_2}dE\,F(\theta,E)\frac{d\Phi}{dE}(E)}{\int_{E_1}^{E_2}dE\,\frac{d\Phi}{dE}(E)}.
\end{equation}
The 68\% containment angle of this cumulative PSF is determined numerically via
\begin{equation}
  0.68=2\upi\int_0^{\theta_{68}}d\theta\,\sin\theta F(\theta,E_1,E_2)
\end{equation}
for a power law spectrum with slope -2.2, the median spectral slope of the point sources in the 3FGL catalog. This is converted to beam width $\sigma_b$, and the cumulative PSF is approximated as a Fisher distribution with the same value of $\sigma_b$. The resulting effective PSF widths are provided for each energy bin in Table~\ref{tab:model}.

\begin{table}
  \caption{The modeled width $\sigma_b$ of the point spread function of \emph{Fermi}-LAT, the modeled flux of the isotropic gamma-ray background for each of our energy bins $E_1\leq E\leq E_2$.}
  \label{tab:model}
  \begin{tabular}{ccccc}
  \hline
  $E_1$ & $E_2$ & $\theta_{68}$ & $\sigma_b$ & $d\Phi_{\text{IGRB}}/d\Omega$ \\
  (GeV) & (GeV) & (${}^\circ$) & ($10^{-3}$ rad) & ($10^{-6}\text{ cm}^{-2}\text{s}^{-1}\text{sr}^{-1}$) \\
  \hline
  2 & 4 & 0.38 & 4.4 & 0.21 \\
  4 & 10 & 0.22 & 2.6 & 0.082 \\
  10 & 20 & 0.14 & 1.6 & 0.017 \\
  20 & 300 & 0.11 & 1.3 & 0.0071 \\
  \hline
  \end{tabular}
\end{table}

\begin{figure}
  \includegraphics[width=\columnwidth]{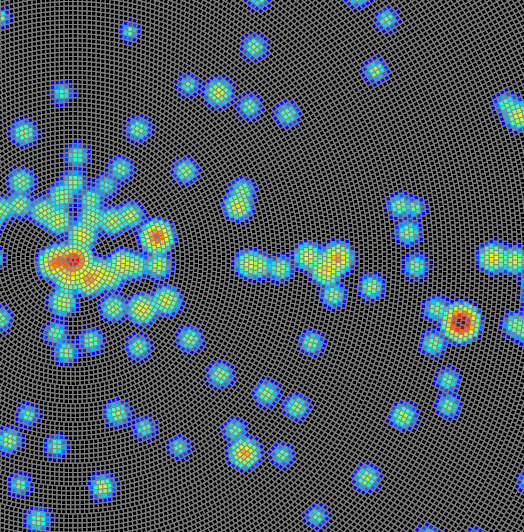}
  \caption{\label{fig:3fgl} A portion of our model of the 3FGL catalog for a $0\fdg2$ GI tiling.}
\end{figure}

Having approximated the effective PSF for each energy bin, we can pixelize the 3FGL catalog as follows. Consider a point source at radial position $\psi_0$ and azimuth $\alpha_0$. The fraction $f_\Phi$ of that source's flux that is observed in a pixel with radial range $\psi_1<\psi<\psi_2$ and azimuth range $\alpha_1<\alpha<\alpha_2$ is
\begin{equation}
  f_\Phi\simeq\frac{1}{\upi\sigma_b^2}\int_{\psi_1}^{\psi_2}\!\!d\psi\,\sin\psi\ I\!\left(\psi\!-\!\psi_0,\psi\!+\!\psi_0;\frac{\alpha_1\!-\!\alpha_0}{2},\frac{\alpha_2\!-\!\alpha_0}{2}\right)
\end{equation}
where
\begin{equation}
  I\!(\varepsilon, \Sigma;\beta_1,\beta_2)\!=\!\!\int_{\beta_1}^{\beta_2}\!\!\!d\beta\,\exp\!\!\left[\!\frac{(\cos\Sigma\!-\!\cos\varepsilon)\sin^2\!\beta\!-\!2\sin^2\!\left(\frac{\varepsilon}{2}\right)}{\sigma_b^2}\!\right].
\end{equation}
A sample of our model of the 3FGL catalog at $0\fdg2$ pixel resolution is shown in \fig{fig:3fgl}.

\section{Isotropic $\gamma$-Ray Background Model}
\label{app:igrb}

\begin{figure}
  \includegraphics[width=\columnwidth]{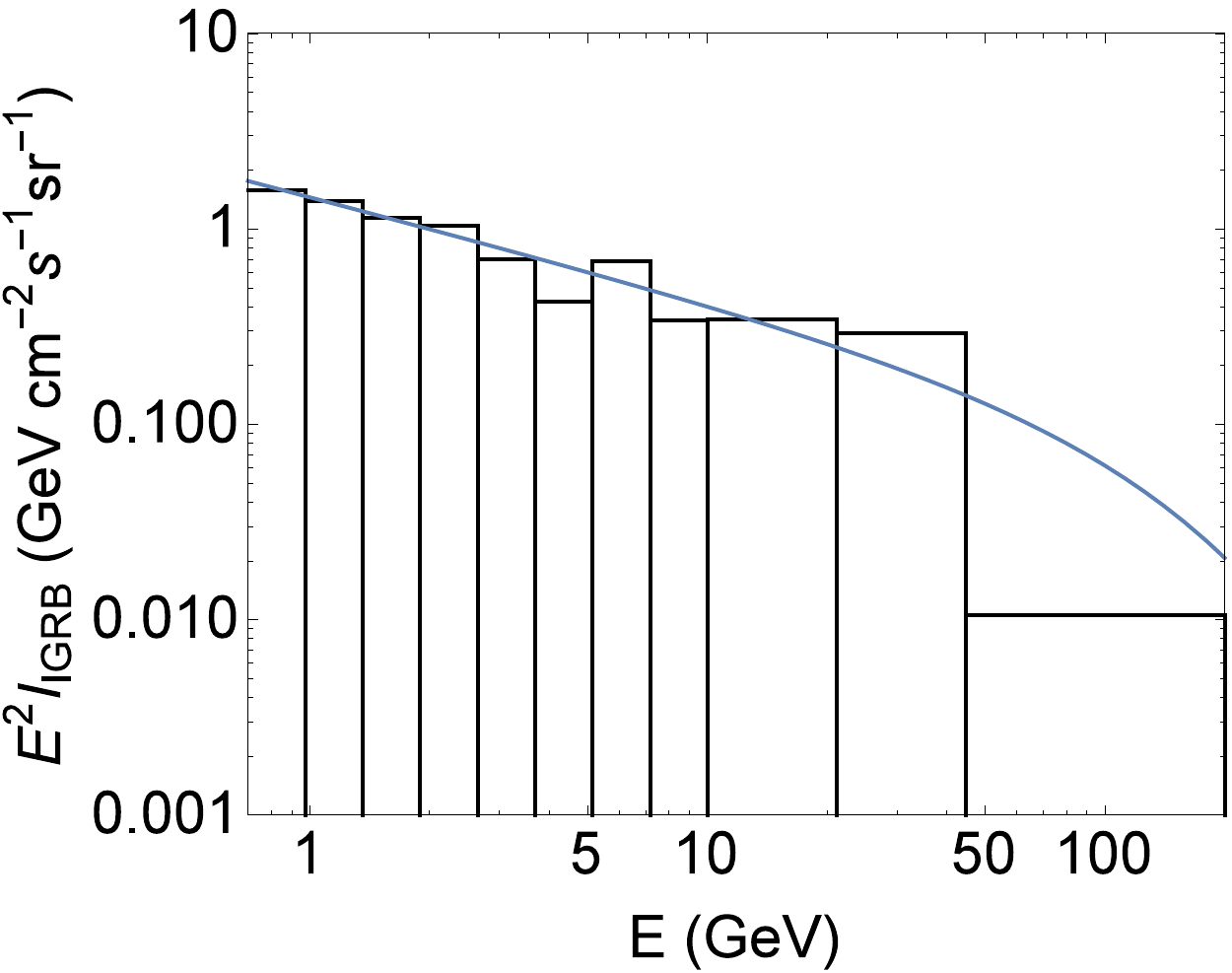}
  \caption{\label{fig:igrb}The data for the best-fit IGRB spectrum for our foreground model, and the power-law with exponential cutoff fit to that data.}
\end{figure}

The best-fit spectrum of the isotropic $\gamma$-ray background (IGRB) is determined for our foreground model in \citet{Horiuchi:2016zwu}, determined as a histogram in $E^2 d\Phi/dE$ and shown in \fig{fig:igrb}. We approximate the IGRB flux in each of our energy bins by fitting a power-law spectrum with exponential cutoff to the histogram and integrating the resulting spectrum over each energy bin.

The spectrum is fit using a goodness-of-fit test that gives higher weight to more precisely determined bins via an inverse-variance weighting. The model intensity spectrum is a power law with exponential cutoff
\begin{equation}
  I(E|I_0,\beta,E_c)=I_0\left(\frac{E}{0.1 \text{GeV}}\right)^{\!\!-\beta}e^{-E/E_c}.
\end{equation}
We defined the best fit by minimizing the $\chi^2$ statistic
\begin{equation}
  \chi^2(I_0,\beta,E_c)=\sum_i\frac{[D_i-M_i(I_0,\beta,E_c)]^2}{\sigma_i^2(I_0,\beta,E_c)},
\end{equation}
where the index $i$ iterates over the energy bins of the data, each with energy range $E_{1,i}\leq E\leq E_{2,i}$. The data values $D_i$ are the specified average of $E^2I(E)$ in each energy bin, and the modeled values are
\begin{equation}
  M_i=\frac{1}{E_{2,i}-E_{1,i}}\int_{E_{1,i}}^{E_{2,i}}dE\,E^2\,I\!(E).
\end{equation}

The variance $\sigma_i^2$ is estimated as the Poisson noise for observing the model. This variance is understood as follows.

For a given energy bin $i$ with width $\Delta E$ and an ensemble of $N$ observed events with energies $E_1,\ldots,E_N$ after exposure $\varepsilon$, the observed data is
\begin{equation}
  \overline{E^2I(E)}=\frac{1}{\varepsilon\Delta E}\sum_{j=1}^NE_j^2.
\end{equation}

The IGRB flux in the energy bin is
\begin{equation}
  \frac{d\Phi}{d\Omega}(\Delta E)=\int_{\Delta E}dE\,I\!(E).
\end{equation}
The probability distribution function of each event's energy is proportional to the model spectrum $I(E)$. The probability distribution of the number of observed events $N$ is just the Poisson distribution \eqn{eqn:poisson} with $n_{\text{exp}}=\varepsilon d\Phi/d\Omega$ for each energy bin's exposure and flux. An ensemble average $\mean{\cdot}$ over event counts and energies involves modularizing over the energy of each event, and then over the number of events.

Then the Poisson variance of $E^2I$ in an energy bin is
\begin{align}
  &\sigma^2=\mean{\overline{E^2I}^2}-\mean{\overline{E^2I}}^{\!\!2}\nonumber\\
  &=\sum_{N=0}^\infty P_P\Big(N\Big|\varepsilon\frac{d\Phi}{d\Omega}\Big)\int_{\Delta E}\left[\prod_{k=1}^NdE_k\frac{I(E_k)}{d\Phi/d\Omega}\right]\left(\sum_{j=1}^N\frac{E_j^2}{\varepsilon\Delta E}\right)^{\!\!2}\nonumber\\
  &\quad-\left\{\sum_{N=0}^\infty P_P\Big(N\Big|\varepsilon\frac{d\Phi}{d\Omega}\Big)\int_{\Delta E}\left[\prod_{k=1}^NdE_k\frac{I(E_k)}{d\Phi/d\Omega}\right]\sum_{j=1}^N\frac{E_j^2}{\varepsilon\Delta E}\right\}^2\nonumber\\
  &=\frac{1}{\varepsilon^2(\Delta E)^2d\Phi/d\Omega}\sum_{N=0}^\infty P_P(N)N\int_{\Delta E}dE\,E^4I(E)\nonumber\\
  &\quad+\frac{1}{(\varepsilon\Delta E d\Phi/d\Omega)^2}\sum_{N=0}^\infty P_P(N)N(N-1)\left[\int_{\Delta E}\!\!\!dE\,E^2I(E)\right]^2\nonumber\\
  &\quad-\left\{\frac{1}{\varepsilon\Delta E d\Phi/d\Omega}\sum_{N=0}^\infty P_P(N)N\int_{\Delta E}dE\,E^2I(E)\right\}^2\nonumber\\
  &=\frac{1}{\varepsilon(\Delta E)^2}\int_{\Delta E}dE\,E^4I(E),
\end{align}
where the last line made use of the Poisson moments
\begin{align}
  &\sum_NNP_P(N|n_{\text{exp}})=n_{\text{exp}},\\
  &\sum_NN(N-1)P_P(N|n_{\text{exp}})=n_{\text{exp}}^2.
\end{align}

This method results in an IGRB spectrum with $I_0=4.97\times10^{-4}$ GeV$^{-1}$cm$^{-2}$s$^{-1}$sr$^{-1}$, $\beta=2.53$, and $E_c=139$ GeV, and is shown by the blue curve in \fig{fig:igrb}. The resulting IGRB fluxes in each of the four energy bins of our GI flux analysis are provided in Table~\ref{tab:model}.

\bsp
\label{lastpage}
\end{document}